\documentclass[12pt, draftclsnofoot, onecolumn]{IEEEtran}

\usepackage{amssymb}
\usepackage{amsmath}
\usepackage{cite}
\usepackage{url}
\usepackage{empheq}
\usepackage{xcolor}
\usepackage{graphicx}
\usepackage{subfigure}
\usepackage{enumitem}
\usepackage{fancyhdr}
\usepackage{mdwmath}	
\usepackage{mdwtab}
\usepackage{caption}
\usepackage{amsthm}
\usepackage{algorithm}
\usepackage{algorithmic}

\newtheorem{lemma}{Lemma}

\newtheorem{theorem}{Theorem}
\newtheorem{corollary}{Corollary}
\newtheorem{assumption}{Assumption}
\newtheorem{definition}{Definition}
\newtheorem{proposition}{Proposition}

\newcommand{\eqr}[1]{(\ref{#1})}
\newcommand{\fref}[1]{Fig.~\ref{#1}}

\newcommand*{\QEDA}{\null\nobreak\hfill\ensuremath{\blacksquare}}

\DeclareMathOperator*{\argmax}{argmax}

\begin{document}
\title{Antenna Activation and Resource Allocation in Multi-Waveguide Pinching-Antenna Systems}
\author{Kaidi~Wang,~\IEEEmembership{Member,~IEEE,}
Zhiguo~Ding,~\IEEEmembership{Fellow, IEEE,}
and George~K.~Karagiannidis,~\IEEEmembership{Fellow, IEEE}
\thanks{K. Wang and Z. Ding are with the Department of Electrical and Electronic Engineering, the University of Manchester, M1 9BB Manchester, UK (email: kaidi.wang@ieee.org; zhiguo.ding@ieee.org).}
\thanks{Z. Ding is also with Khalifa University, Abu Dhabi, UAE.}
\thanks{G. K. Karagiannidis is with Department of Electrical and Computer Engineering, Aristotle University of Thessaloniki, Greece (email: geokarag@auth.gr).}}
\maketitle
%%%%%%%%%%%%%%%%%%%%%%%%%%%%%%%%%%%%%%%%%%%%%%%%%
%%%%%%%%%%%%%%%%%%%%%%%%%%%%%%%%%%%%%%%%%%%%%%%%%
\begin{abstract}
Pinching antennas, as a novel flexible-antenna technology capable of establishing line of sight (LoS) connections and effectively mitigating large-scale path loss, have recently attracted considerable research interests. However, the implementation of ideal pinching-antenna systems involves determining and adjusting pinching antennas to an arbitrary position on waveguides, which presents challenges to both practical deployment and related optimization. This paper investigates a practical pinching-antennas system in multi-waveguide scenarios, where pinching antennas are installed at pre-configured discrete positions to serve downlink users with non-orthogonal multiple access (NOMA). To improve system throughput, a sophisticated optimization problem is formulated by jointly considering waveguide assignment, antenna activation, successive interference cancellation (SIC) decoding order design, and power allocation. By treating waveguide assignment and antenna activation as two coalition-formation games, a novel game-theoretic algorithm is developed, in which the optimal decoding order is derived and incorporated. For power allocation, monotonic optimization and successive convex approximation (SCA) are employed to construct global optimal and low-complexity solutions, respectively. Simulation results demonstrate that the NOMA-based pinching-antenna system exhibits superior performance compared to the considered benchmark systems, and the proposed solutions provide significant improvement in terms of sum rate and outage probability.
\end{abstract}
%%%%%%%%%%%%%%%%%%%%%%%%%%%%%%%%%%%%%%%%%%%%%%%%% 
%%%%%%%%%%%%%%%%%%%%%%%%%%%%%%%%%%%%%%%%%%%%%%%%%
\section{Introduction}
With the ongoing development of sixth-generation (6G) communications, various advanced technologies have been introduced to improve the throughput and reliability of wireless communications. Due to their ability to dynamically reconfigure effective wireless channels, flexible-antenna techniques, such as intelligent reflecting surfaces (IRSs) \cite{wu2019irs}, fluid antennas \cite{wong2020fluid}, and movable antennas \cite{zhu2023modeling}, have gained significant attention as promising research directions in both academia and industry. Specifically, IRSs focus on using reflective elements to redirect signals in desired directions \cite{wu2021intelligent}, while fluid and movable antennas aim to physically change antenna positions \cite{wu2024fluid, ma2024movable}. Both analytical and optimization-based studies have demonstrated the superiority of these flexible-antenna systems over conventional fixed-antenna configurations.

While introducing notable performance improvements, existing flexible-antenna technologies also exhibit certain limitations, primarily reflected in their ineffectiveness in mitigating large-scale path loss \cite{yang2025pinching}. For example, IRSs rely on the adjustment of phase shifts to construct a reflected line of sight (LoS) between the transceiver, which inherently results in double attenuation. Similarly, fluid and movable antennas can only adjust antenna positions within a few wavelengths, resulting in relatively small effects, especially in scenarios where the LoS link is obstructed. In this context, pinching antennas have recently been proposed as a novel and promising evolution of the family of flexible-antenna technologies \cite{suzuki2022pinching, ding2024pin}. By applying small dielectric particles along a dielectric waveguide, the wireless channel can be dynamically reconfigured \cite{liu2025pinching}. Because of their ability to adjust spatially over large areas, pinching-antenna systems offer significant advantages over conventional flexible-antenna systems, particularly in combating large-scale path loss and establishing LoS links.
%%%%%%%%%%%%%%%%%%%%%%%%%%%%%%%%%%%%%%%%%%%%%%%%%
\subsection{Related Works}
In order to explore the achievable performance of pinching-antenna systems, many research efforts have recently conducted from the analysis and optimization perspectives \cite{ouyang2025array, xu2025pin, tyrovolas2025pin, qin2025joint, mu2025pin, bereyhi2025pin}. In \cite{ouyang2025array}, the upper bound on the array gain for pinching-antenna systems was analyzed, emphasizing the importance and necessity of optimizing the number of activated pinching antennas and the antenna spacing. To optimize the achievable data rate in a single-user pinching-antenna system, a challenging antenna scheduling problem was formulated and transformed into a low-complexity problem to minimize the impact of large-scale path loss \cite{xu2025pin}. By incorporating the waveguide losses, the authors of \cite{tyrovolas2025pin} analyzed the expressions of the outage probability and the average rate in a particle pinching-antenna system. The optimal placement of pinching antennas was determined to enhance the system performance while mitigating the effect of waveguide attenuation. Pinching antennas were also investigated in integrated sensing and communication (ISAC) systems to enhance the sum rate, which was achieved by jointly solving antenna scheduling and power allocation problems using a reinforcement learning based algorithm \cite{qin2025joint}. In \cite{mu2025pin}, a minimum data rate maximization problem was investigated in a pinching-antenna system enabled multicast communications scenario, where a particle swarm optimization based algorithm was developed to determine the optimal positions of pinching antennas. The hybrid beamforming design for pinching-antenna systems was explored in \cite{bereyhi2025pin}, where multiple single-antenna waveguides were implemented to simultaneously serve users. Through the optimization of the digital precoding and analogue precoding (via changing antenna locations), the weighted sum rate of the considered system was maximized.

One characteristic of pinch-antenna systems is that all antennas located on the same waveguide must transmit the same information, and then non-orthogonal multiple access (NOMA) becomes a potential transmission solution for multi-user pinching-antenna systems \cite{ding2025blockage, xie2025pin, hu2025pin, kaidi2025pin}. In \cite{ding2025blockage}, the capability of pinching-antenna systems for mitigating blockages and establishing LoS links was demonstrated. It was underscored that pinching-antenna systems are particularly advantageous for NOMA schemes, as the co-channel interference can be effectively suppressed by LoS blockage. The authors of \cite{xie2025pin} investigated pinching-antenna systems in both orthogonal multiple access (OMA) and NOMA scenarios. By employing an offset that represents the distance between the pinching antenna and the closest position of each user on the waveguide, a low-complexity antenna scheduling algorithm was proposed to maximize the sum rate. Furthermore, the NOMA-assisted pinching-antenna system was also studied in multi-waveguide scenarios, where the beamforming design and antenna placement were jointly optimized with the objective to maximize the sum rate \cite{hu2025pin}. To achieve this objective, two successive convex approximation (SCA) based algorithms were developed and implemented iteratively using alternating optimization. In contrast to the conventional optimization of antenna placement, \cite{kaidi2025pin} proposed a low-complexity implementation of pinching-antenna systems by considering pre-configured pinching antennas. Based on antenna activation, a sum rate maximization problem was formulated to enhance the system performance.
%%%%%%%%%%%%%%%%%%%%%%%%%%%%%%%%%%%%%%%%%%%%%%%%%
\subsection{Motivation and Contributions}
In contrast to the aforementioned works \cite{ouyang2025array, xu2025pin, tyrovolas2025pin, qin2025joint, mu2025pin, bereyhi2025pin, ding2025blockage, xie2025pin, hu2025pin} on pinching-antenna systems, in this work, a practical implementation of pinching-antenna systems is considered, where pinching antennas are installed at discrete pre-configured positions along the waveguide and are dynamically activated/deactivated to serve users. Consequently, the delay caused by the physical movement of the pinching antennas on the waveguide can be mitigated, and the high-complexity antenna placement problem can be simplified and replaced by a more tractable antenna activation problem. Compared to \cite{kaidi2025pin}, which focuses only on a single-waveguide scenario, this work studies a multi-waveguide setting. This extension enhances overall system performance but also introduces new challenges, such as the assignment of users among waveguides and the impact of inter-waveguide interference on successive interference cancellation (SIC) decoding. In addition, power allocation is incorporated to further enhance the performance of the NOMA-based system. The main contributions of this paper are summarized as follows:
\begin{itemize}
\item A practical multi-waveguide pinching-antenna system is designed, in which pinching antennas are evenly distributed at pre-configured discrete positions along the waveguides. By activating or deactivating these pre-configured antennas, a low-complexity antenna scheduling strategy is achieved. In addition, through the adoption of NOMA transmission schemes, the pinching antennas on each waveguide can simultaneously serve multiple associated users with different power levels.
\item To further enhance system performance, a sum rate maximization problem under quality of service (QoS) constraints is formulated by jointly considering waveguide assignment, antenna activation, power allocation, and SIC decoding order design. Due to the fact that there are both discrete and continuous optimization variables, the resulting mixed-integer optimization problem is decomposed into two subproblems, including a waveguide assignment and antenna activation problem and a power allocation problem, and solved in an alternative manner.
\item The optimal SIC decoding order is first determined in the proposed pinching-antenna system, ensuring successful implementation of SIC without degrading the achievable data rate of any user. Then, waveguide assignment and antenna activation are modeled as two coalition-formation games in partition form. Based on this framework, a game-theoretic algorithm is developed that is capable of iteratively incorporating power allocation and SIC decoding order design. The properties and performance of the proposed algorithm are also analyzed.
\item The power allocation problem is solved through two different approaches: a globally optimal solution and a computationally efficient alternative. Specifically, based on monotonic optimization, a polyblock outer approximation algorithm is proposed to approach the global optimum. In contrast, the original problem is transformed into a convex approximation and a low-complexity solution is obtained using a SCA based algorithm.
\end{itemize}
In simulation results, it is demonstrated that the proposed NOMA-based multi-waveguide pinching-antenna system is able to outperform both the OMA-based pinching-antenna system and the NOMA-based single-waveguide pinching-antenna system. Moreover, the proposed solutions, including the optimal SIC decoding order, the coalitional game based waveguide assignment and antenna activation algorithm, and the power allocation strategies, can provide significant improvements in terms of both sum rate and outage probability.
%%%%%%%%%%%%%%%%%%%%%%%%%%%%%%%%%%%%%%%%%%%%%%%%%
%\subsection{Organization}
%The remainder of this paper is organized as follows. Sections II and III present the system model of the pinching-antenna architecture and the formulation of the sum rate maximization problem, respectively. The proposed solutions for decoding order design, waveguide assignment, antenna activation, and power allocation are detailed in Sections IV and V. Simulation results are provided in Section VI, and the conclusions are summarized in Section VII.
%%%%%%%%%%%%%%%%%%%%%%%%%%%%%%%%%%%%%%%%%%%%%%%%%
%%%%%%%%%%%%%%%%%%%%%%%%%%%%%%%%%%%%%%%%%%%%%%%%%
\section{System Model}
\begin{figure}[!t]
\centering{\includegraphics[width=85mm]{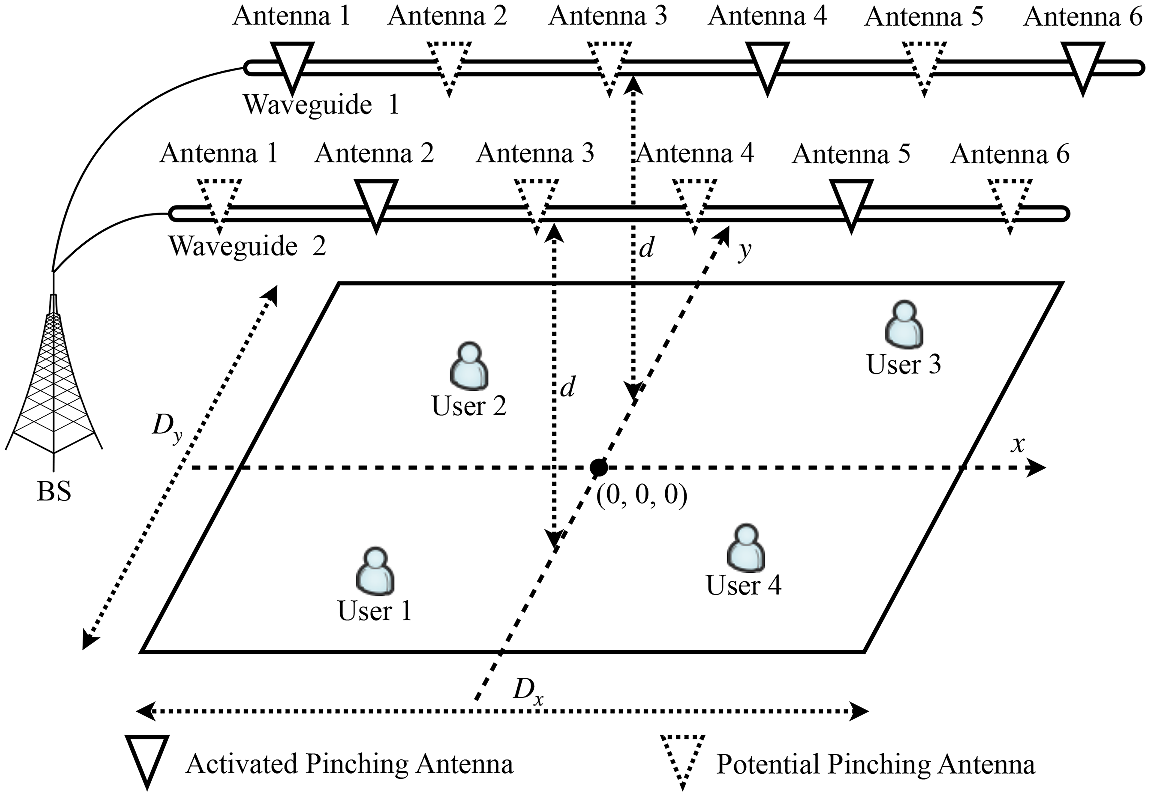}}
\caption{An illustration of the considered NOMA-assisted pinching-antenna system with $K=2$ waveguides, $N=4$ users, and $M=6$ potential pinching antennas on each waveguide.}
\label{system}
\end{figure}

Consider a downlink pinching-antenna system with one base station (BS), $K$ waveguides, and $N$ single-antenna users, where the waveguides are deployed at height $d$ and the users are randomly distributed within a rectangular area with side lengths $D_x$ and $D_y$, as shown in \fref{system}. The collections of all waveguides and users are denoted by $\mathcal{K}=\{1,2,\cdots,K\}$ and $\mathcal{N}=\{1,2,\cdots,N\}$, respectively. On each waveguide, $M$ potential pinching antennas are evenly distributed with antenna spacing $D_x/(M-1)$, and the collection of all potential pinching antennas on each waveguide is denoted by $\mathcal{M}=\{1,2,\cdots,M\}$.

In the considered system, the locations of user $n$ and pinching antenna $m$ on waveguide $k$ are represented by $\boldsymbol{\psi}_n=\left(x_n,y_n,0\right)$ and $\boldsymbol{\psi}_{k,m}^\mathrm{Pin}=\left(x_m^\mathrm{Pin},y_k^\mathrm{Pin},d\right)$, respectively. Based on the spherical wave channel model, in the case that all pinching antennas are activated, the channel vector between waveguide $k$ and user $n$ is given by
\begin{equation}
\bar{\mathbf{h}}_{k,n} = \!\!\left[\frac{\eta e^{-j\frac{2\pi}{\lambda}\left|\boldsymbol{\psi}_n\!-\boldsymbol{\psi}_{k,1}^\mathrm{Pin}\right|}}{\left|\boldsymbol{\psi}_n-\boldsymbol{\psi}_{k,1}^\mathrm{Pin}\right|} \quad \!\!\cdots\!\! \quad \frac{\eta e^{-j\frac{2\pi}{\lambda}\left|\boldsymbol{\psi}_n\!-\boldsymbol{\psi}_{k,M}^\mathrm{Pin}\right|}}{\left|\boldsymbol{\psi}_n-\boldsymbol{\psi}_{k,M}^\mathrm{Pin}\right|}\right]^T,
\end{equation}
where $\eta=\frac{c}{4\pi f_c}$ is the frequency dependent factor, $c$ is the speed of light, $f_c$ is the carrier frequency, and $\lambda$ is the wavelength. In particular, $\left|\boldsymbol{\psi}_n-\boldsymbol{\psi}_{k,m}^\mathrm{Pin}\right|$ indicates the distance between user $n$ and pinching antenna $m$ on waveguide $k$, i.e.,
\begin{equation}
\left|\boldsymbol{\psi}_n-\boldsymbol{\psi}_{k,m}^\mathrm{Pin}\right|=\sqrt{(x_n\!-x_m^\mathrm{Pin})^2\!+\!(y_n\!-y_k^\mathrm{Pin})^2\!+\!d^2}.
\end{equation}

In order to compensate the phase shifts caused by the signal propagation inside each waveguide \cite{pozar2021microwave}, the signal vector transmitted through waveguide $k$ is processed as follows:
\begin{equation}
\bar{\mathbf{x}}_k\!\!=\!\sqrt{\frac{P_t}{M}}\!\!\left[e^{-j\frac{2\pi}{\lambda_g}\left|\boldsymbol{\psi}_{k,0}^\mathrm{Pin}\!-\boldsymbol{\psi}_{k,1}^\mathrm{Pin}\right|} \!\quad\! \!\!\cdots\!\!\! \!\quad\! e^{-j\frac{2\pi}{\lambda_g}\left|\boldsymbol{\psi}_{k,0}^\mathrm{Pin}\!-\boldsymbol{\psi}_{k,M}^\mathrm{Pin}\right|}\right]^T\!\!\!\bar{s}_k,
\end{equation}
where $P_t$ is the available transmit power at each waveguide, $\lambda_g=\lambda/n_\mathrm{eff}$ is the guided wavelength, $n_\mathrm{eff}$ is the effective refractive index of the dielectric waveguide \cite{pozar2021microwave}, $\boldsymbol{\psi}_{k,0}^\mathrm{Pin}=\left(x_0^\mathrm{Pin},y_k^\mathrm{Pin},d\right)$ is the location of the feed point of waveguide $k$, and $\bar{s}_k$ is the transmitted signal at waveguide $k$. 

%%%%%%%%%%%%%%%%%%%%%%%%%%%%%%%%%%%%%%%%%%%%%%%%%
\subsection{Waveguide Assignment and Antenna Activation}
Recall that in pinching-antenna systems, the signal fed into the same waveguide for all antennas must be the same \cite{ding2024pin}. In this paper, it is considered that each user is assigned to one waveguide, and each waveguide transmits the superimposed signal of all assigned users simultaneously, i.e., NOMA transmission schemes. By defining a waveguide assignment indicator $\alpha_{k,n}$, the signal transmitted through the pinching antennas on waveguide $k$ can be expressed as follows:
\begin{equation}
\bar{s}_k=\sum_{n=1}^N\sqrt{p_{k,n}}\alpha_{k,n}s_n,
\end{equation}
where $p_{k,n}\in [0,1]$ is the power allocation coefficient of user $n$ assigned to waveguide $k$, and $s_n$ is the desired signal of user $n$. Specifically, $\sum_{n=1}^N\alpha_{k,n}p_{k,n}\le 1$ and $\alpha_{k,n}\in\{0,1\}$, where $\alpha_{k,n}=1$ indicates that user $n$ is assigned to waveguide $k$, otherwise $\alpha_{k,n}=0$. It is worth mentioning that the number of users assigned to each waveguide is unlimited, and any waveguide can be idle, that is, not serving users.

In addition to waveguide assignment, antenna activation is also considered to yield a practical implementation of pinching-antenna systems. That is, on each waveguide, only a subset of pinching antennas are activated from all potential antennas. To this end, antenna activation indicator $\beta_{k,m}\in\{0,1\}$ is defined to indicate the state of pinching antenna $m$ on waveguide $k$, and the antenna activation indicators of waveguide $k$ can be presented as a vector, i.e., $\boldsymbol{\beta}_k=[\beta_{k,1} \!\quad\! \beta_{k,2} \!\quad\!\cdots\!\quad\! \beta_{k,M}]^T$. Therefore, with the given antenna activation indicators, the channel vector between waveguide $k$ and user $n$ can be expressed as follows:
\begin{equation}
\mathbf{h}_{k,n} = \mathbf{B}_k\bar{\mathbf{h}}_{k,n},
\end{equation}
where $\mathbf{B}_k = diag(\boldsymbol{\beta}_k)$ is a $M\times M$ diagonal matrix. Moreover, the signal vector transmitted through waveguide $k$ can be rewritten as follows:
\begin{equation}
\mathbf{x}_k=\sqrt{\frac{M}{M_k}}\mathbf{B}_k\bar{\mathbf{x}}_k,
\end{equation}
where $M_k$ is the number of all activated antennas on waveguide $k$. It is worth noting that $\frac{M}{M_k}$ is included in the above equation to ensure that the transmit power is equally allocated among all activated pinching antennas.

%%%%%%%%%%%%%%%%%%%%%%%%%%%%%%%%%%%%%%%%%%%%%%%%%
\subsection{NOMA-based Signal Model}
In the considered pinching-antenna system, the NOMA principle is employed, and hence, the signals transmitted from all waveguides can be received at all users. The received signal at any user $n$ is given by
\begin{equation}
y_n=\sum_{k=1}^K\mathbf{h}_{k,n}^H\mathbf{x}_k\!+\!w_n=\sum_{k=1}^Kh_{k,n}\!\sum_{i=1}^N\alpha_{k,i}\sqrt{p_{k,i}\frac{P_t}{M_k}}s_i\!+\!w_n,
\end{equation}
where
\begin{equation}
h_{k,n}\triangleq\sum_{m=1}^M\beta_{k,m}\frac{\eta e^{-2\pi j\left(\frac{1}{\lambda}\left|\boldsymbol{\psi}_n\!-\boldsymbol{\psi}_{k,m}^\mathrm{Pin}\right|+\!\frac{1}{\lambda_g}\left|\boldsymbol{\psi}_{k,0}^\mathrm{Pin}\!-\boldsymbol{\psi}_{k,m}^\mathrm{Pin}\right|\right)}}{\left|\boldsymbol{\psi}_n-\boldsymbol{\psi}_{k,m}^\mathrm{Pin}\right|},
\end{equation}
and $w_n$ is the additive white Gaussian noise. Assuming that user $n$ is assigned to waveguide $k$, i.e., $\alpha_{k,n}=1$, the received signal at user $n$ can be rewritten as follows:
\begin{align}
y_n =\underbrace{h_{k,n}\sqrt{p_{k,n}\frac{P_t}{M_k}}s_n}_{\text{desired signal}}+\underbrace{h_{k,n}\sum_{i=1}^N\alpha_{k,i}\sqrt{p_{k,i}\frac{P_t}{M_k}}s_i}_{\text{intra-waveguide interference}}+\underbrace{\sum_{k'=1,k'\neq k}^K\!\!\!\!h_{k',n}\sum_{i=1}^N\alpha_{k',i}\sqrt{p_{k',i}\frac{P_t}{M_{k'}}}s_i}_{\text{inter-waveguide interference}}+\omega_n.
\end{align}

It is worth noting that in the considered system, the intra-waveguide interference can be partially removed by utilizing SIC techniques. That is, with a given decoding order, each user can first decode and remove the signals for the users assigned to the same waveguide and preceding it in the decoding order, and then decode the desired signal by treating the signals for other users as noise \cite{ding2014noma}. In the case that the decoding order on waveguide $k$ (denoted by $\mathrm{Q}_k$) is from user $1$ to user $N$, the data rate for user $n$ decoding its own signal is given by
\begin{equation}\label{rate}
R_{k,n\to n}^{(\mathrm{Q}_k)}\!=\log_2\!\!\left(\!1\!+\!\frac{p_{k,n}\frac{P_t}{M_k}|h_{k,n}|^2}{\frac{P_t}{M_k}|h_{k,n}|^2\!\sum_{i=n+1}^N\!\alpha_{k,i}p_{k,i}\!+\!I_{k,n}\!+\!\sigma^2}\!\right),
\end{equation}
where
\begin{equation}\label{interference}
I_{k,n} = \!\!\sum_{k'=1,k'\neq k}^K\!\!\frac{P_t}{M_{k'}}|h_{k',n}|^2\sum_{i=1}^N\alpha_{k',i}p_{k',i},
\end{equation}
and $\sigma^2$ is the noise power. Accordingly, the desired signal of user $n$ needs to be decoded by the subsequent users assigned to the same waveguide, i.e., $n', \forall n' > n$. The data rate for user $n'$ decoding the desired signal of user $n$ is given by
\begin{equation}\label{decoderate}
R_{k,n'\to n}^{(\mathrm{Q}_k)}=\log_2\!\!\left(\!1\!+\!\frac{p_{k,n}\frac{P_t}{M_k}|h_{k,n'}|^2}{\frac{P_t}{M_k}|h_{k,n'}|^2\!\sum_{i=n+1}^N\!\alpha_{k,i}p_{k,i}\!+\!I_{k,n'}\!+\!\sigma^2}\!\right).
\end{equation}

Note that in order to guarantee that user $n'$ can successfully decode the desired signal of user $n$, the data rate for transmitting the signal of user $n$ should be limited \cite{kaidi2024nfc}. Therefore, the achievable data rate of user $n$ assigned to waveguide $k$ with decoding order $\mathrm{Q}_k$ can be presented as follows:
\begin{equation}\label{achirate}
R_{k,n}^{(\mathrm{Q}_k)}=\min\left\{R_{k,i\to n}^{(\mathrm{Q}_k)} |\forall i\ge n\right\}.
\end{equation}
In conventional downlink NOMA systems, an optimal SIC decoding order can be obtained by sorting users based on their channel conditions, thereby avoiding the situation of reducing the data rate. However, this method is not applicable to the proposed multi-waveguide pinching-antenna system due to the existence of inter-waveguide interference. Therefore, the SIC decoding order is also considered as an optimization parameter in this paper.

%%%%%%%%%%%%%%%%%%%%%%%%%%%%%%%%%%%%%%%%%%%%%%%%%
%%%%%%%%%%%%%%%%%%%%%%%%%%%%%%%%%%%%%%%%%%%%%%%%%
\section{Problem Formulation}
This work focuses on improving the system throughput of the proposed pinching-antenna system, and hence, by jointly considering waveguide assignment, antenna activation, power allocation, and decoding order design, a sum rate maximization problem is formulated as follows:
\begin{subequations}
\begin{empheq}{align}
\max_{\boldsymbol{\alpha},\boldsymbol{\beta},\mathbf{P},\mathbf{Q}}\quad & \sum_{k=1}^K\sum_{n=1}^N\alpha_{k,n}R_{k,n}^{(\mathrm{Q}_k)}\\
\textrm{s.t.} \quad & R_{k,n}^{(\mathrm{Q}_k)}\ge \alpha_{k,n}R_\mathrm{min}, \forall n\in\mathcal{N},\\
& 0\le p_{k,n}, \forall n\in\mathcal{N}, \forall k \in\mathcal{K},\\ 
& \sum\nolimits_{n=1}^N\alpha_{k,n}p_{k,n} \le 1, \forall k \in\mathcal{K},\\
& \alpha_{k,n}\in\{0,1\}, \forall k\in\mathcal{K}, \forall n\in\mathcal{N},\\
& \sum\nolimits_{k=1}^K\alpha_{k,n}=1, \forall n\in\mathcal{N},\\
& \beta_{k,m}\in\{0,1\}, \forall k\in\mathcal{K}, \forall m\in\mathcal{M},
\end{empheq}
\label{problem}
\end{subequations}\vspace{-2mm}\\
where $\boldsymbol{\alpha}$, $\boldsymbol{\beta}$, $\mathbf{P}$, and $\mathbf{Q}$ are the collections of all waveguide assignment indicators, antenna activation indicators, power allocation coefficients, and SIC decoding orders, respectively. In constraint (\ref{problem}b), the QoS thresholds are included for all users, where $R_\mathrm{min}$ is the target rate. Constraint (\ref{problem}f) implies that each user must be assigned to one waveguide. In the case that any waveguide is not allocated to users, all pinching antennas on that waveguide are deactivated.

The formulated sum rate maximization problem in \eqr{problem} is a mixed integer optimization problem and is therefore very challenging to solve jointly. In this paper, it is decoupled into two sub-problems and solved via alternative optimization. By including all integer variables, the problem of waveguide assignment, antenna activation, and decoding order design can be presented as follows:
\begin{subequations}
\begin{empheq}{align}
\max_{\boldsymbol{\alpha},\boldsymbol{\beta},\mathbf{Q}}\quad & \sum_{k=1}^K\sum_{n=1}^N\alpha_{k,n}R_{k,n}^{(\mathrm{Q}_k)}\\\nonumber
\textrm{s.t.} \quad & \text{(\ref{problem}e), (\ref{problem}f), and (\ref{problem}g)}.
\end{empheq}
\label{waaaproblem}
\end{subequations}\vspace{-2mm}\\
We note that the QoS constraint is not included in the above problem because considering this constraint in an integer programming problem would limit the performance of its solution. Correspondingly, we ensure QoS through the power allocation problem as follows:
\begin{subequations}
\begin{empheq}{align}
\max_{\mathbf{P}}\quad & \sum_{k=1}^K\sum_{n=1}^N\alpha_{k,n}R_{k,n}^{(\mathrm{Q}_k)}\\\nonumber
\textrm{s.t.} \quad & \text{(\ref{problem}b), (\ref{problem}c), and (\ref{problem}d)}.
\end{empheq}
\label{paproblem}
\end{subequations}\vspace{-2mm}\\

%%%%%%%%%%%%%%%%%%%%%%%%%%%%%%%%%%%%%%%%%%%%%%%%%
%%%%%%%%%%%%%%%%%%%%%%%%%%%%%%%%%%%%%%%%%%%%%%%%%
\section{Solutions for Waveguide Assignment and Antenna Activation}
This section focuses on solving problem \eqr{waaaproblem}, where given power allocation coefficients are exploited. In the considered NOMA-based pinching-antenna system, the effective channel varies with the waveguide assignment and antenna activation strategies. Specifically, for any user, the channels transmitting the desired signal and the interfering signal may be switched due to waveguide assignment, and these channels are also affected by antenna activation. Therefore, the SIC decoding order design depends on the given solution of waveguide assignment and antenna activation. In this section, the optimal SIC decoding order for each waveguide is first developed and then implemented in the proposed waveguide assignment and antenna activation algorithm.

%%%%%%%%%%%%%%%%%%%%%%%%%%%%%%%%%%%%%%%%%%%%%%%%%
%%%%%%%%%%%%%%%%%%%%%%%%%%%%%%%%%%%%%%%%%%%%%%%%%
\subsection{SIC Decoding Order Design}
Due to the fact that user $n$ assigned to waveguide $k$ may not be the $n$-th decoded user in that waveguide, in order to facilitate optimizing SIC decoding orders, a set of subscripts are introduced to represent the users' positions in the decoding order. Without loss of generality, for any waveguide $k$ with $N_k$ assigned users, the decoding order $Q_k$ is a sequence $(1)\to (2)\to\cdots\to (N_k)$, where $(n)$ indicates the $n$-th decoded user. According to the newly defined subscripts, for each waveguide, the following optimal SIC decoding order can be obtained:
\begin{proposition}\label{order}
In the proposed NOMA-based multi-waveguide pinching-antenna system, the optimal SIC decoding order of any waveguide $k$ is $Q_k^*$ if the following condition is satisfied:
\begin{equation}
\frac{I_{k,(1)}\!+\sigma^2}{|h_{k,(1)}|^2}\ge \frac{I_{k,(2)}\!+\sigma^2}{|h_{k,(2)}|^2}\ge\cdots\ge\frac{I_{k,(N_k)}\!+\sigma^2}{|h_{k,(N_k)}|^2}.
\end{equation}
\begin{IEEEproof}
Refer to Appendix A.
\end{IEEEproof}
\end{proposition}

According to Proposition \ref{order}, the optimal SIC decoding order can be generated for each waveguide based on the given solution of waveguide assignment and antenna activation, which is presented in the next subsection. Proposition \ref{order} indicates that the optimal SIC decoding order is mainly determined by the effective channels and inter-waveguide interference, where the users with better channel conditions and/or weaker interference are able to decode other users' signals. Moreover, it is worth pointing out that the inter-waveguide interference is independent of the power allocation coefficients, as demonstrated in Section \ref{mosolution}. Therefore, the optimal SIC decoding order in Proposition \ref{order} can be obtained under any power allocation and does not change accordingly.

Based on the optimal decoding order, the achievable data rate for each user can be calculated without considering the impact of other users, as follows:
\begin{equation}\label{optrate}
R_{k,(n)}^{(\mathrm{Q}_k^*)}\!=\!\log_2\!\!\left(\!\!1\!\!+\!\!\frac{p_{k,(n)}\frac{P_t}{M_k}|h_{k,(n)}|^2}{\frac{P_t}{M_k}|h_{k,(n)}|^2\!\sum_{i=(n)+1}^{(N_k)}\!p_{k,i}\!+\!I_{k,(n)}\!+\!\sigma^2}\!\!\right).
\end{equation}
For the last decoded user $(N_k)$, the achievable data rate can be presented as follows:
\begin{equation}\label{optrate}
R_{k,(N_k)}^{(\mathrm{Q}_k^*)}\!=\!\log_2\!\!\left(\!\!1\!\!+\!\!\frac{p_{k,(N_k)}\frac{P_t}{M_k}|h_{k,(N_k)}|^2}{I_{k,(N_k)}\!+\!\sigma^2}\!\!\right).
\end{equation}
Note that the above equation is also applicable to the OMA case, where there is only one user assigned to the waveguide, i.e., $N_k=1$.

%%%%%%%%%%%%%%%%%%%%%%%%%%%%%%%%%%%%%%%%%%%%%%%%%
%%%%%%%%%%%%%%%%%%%%%%%%%%%%%%%%%%%%%%%%%%%%%%%%%
\subsection{Waveguide Assignment and Antenna Activation}
In problem \eqr{waaaproblem}, waveguide assignment and antenna activation can be considered as two different coalitional games, where users and pinching antennas can join the coalitions formed by waveguides, respectively. Hence, a coalitional game based algorithm is proposed to jointly solve waveguide assignment and antenna activation.

By treating the users assigned to waveguide $k$ as a coalition of players, waveguide assignment can be viewed as a coalition-formation game $(\mathcal{N}, v, \mathcal{A})$, where $v$ is the coalition value, $\mathcal{A}=\{A_1, A_2, \cdots, A_K\}$ is the coalition structure, and $A_k$ is the coalition formed by waveguide $k$. According to constraints (\ref{problem}e) and (\ref{problem}f), it can be obtained that $A_k\cap A_k' =\emptyset, \forall k\neq k'$ and $\cup_{k\in\mathcal{K}}A_k=\mathcal{N}$. For solving the waveguide assignment game, the preferences of users are considered and constructed according to the coalition value, which is defined by the sum rate. Specifically, any player can split from one coalition and then merge into another coalition if the coalition value can be strictly increased. Due to the fact that the inter-waveguide interference changes if a new empty or non-empty coalition is generated, i.e., the only user assigned to any waveguide decides to switch to another waveguide or any user moves to a waveguide with no assigned users, the decision of one user may affect the data rates of other users in all coalitions. This case corresponds to the coalition-formation game in partition form \cite{han2012game}, where the coalition structure plays an important role in determining coalition value. As a result, the coalition value should take into account the coalition structure and include the data rates of all users, as follows:
\begin{equation}
v(\mathcal{A}) = \sum_{A_k\in\mathcal{A}}\sum_{i\in A_k}R_{k,i}^{(\mathrm{Q}_k)}.
\end{equation}
Assuming that user $n$ in coalition $A_k$, where $A_k\in\mathcal{A}$, tends to be assigned to another waveguide $k'$, where $k'\neq k$, the coalition structure is transformed from $\mathcal{A}$ to $\mathcal{A}'$, where $\mathcal{A}'=\mathcal{A}\backslash\{A_k,A_{k'}\}\cup\{A_k\backslash\{n\},A_{k'}\cup\{n\}\}$. This operation follows the strict preference relation, as shown below:\begin{equation}\label{userpref}
(A_k,\mathcal{A}) \!\prec_n\! (A_{k'},\mathcal{A}') \Leftrightarrow v(\mathcal{A})\!<\!v(\mathcal{A}').
\end{equation}
Since the players of the involved coalitions, i.e., $A_k$ and $A_{k'}$, have been changed, the SIC decoding orders and power allocation of both coalitions should be readjusted during this process. To obtain the solution of the waveguide assignment game, Nash stability \cite{bogomolnaia2002stability} is employed, as shown below:
\begin{definition}\label{stability1}
In the waveguide assignment game, a coalition structure $\mathcal{A}$ is Nash stable if there is no user willing to be assigned to another waveguide, as follows:
\begin{equation}
(A_k,\mathcal{A})\succeq_n (A_{k'},\mathcal{A}'), \forall n \in A_k, \forall A_k \in \mathcal{A}, \forall A_{k'} \in \mathcal{A}'.
\end{equation}
\end{definition}

The antenna activation problem can also be viewed as a coalition-formation game $(\mathcal{L}, u, \mathcal{B})$ in partition form, where $\mathcal{L}$ is the collection of all pinching antennas, $u$ is the coalition value, and $\mathcal{B}=\{B_1, B_2,\cdots, B_K\}$ is the partition or the coalition structure. Unlike the waveguide assignment game, in the antenna activation game, players, i.e., pinching antennas, can only join (activation) or leave (deactivation) the coalitions formed by corresponding waveguides. Due to the fact that the activation or deactivation of any pinching antenna on any waveguide changes the data rates of all users through channel vectors, the coalition value is defined as follows:
\begin{equation}
u(\mathcal{B}) = \sum_{k=1}^K\sum_{n=1}^NR_{k,n}^{(\mathrm{Q}_k)}.
\end{equation}
Based on the expression of the coalition value, the preference relations of players can be constructed. For deactivated antenna $m$ on waveguide $k$, it is willing to be activated if the coalition value can be strictly increased, as follows:
\begin{equation}\label{antennaact}
(B_k, \mathcal{B}) \!\prec_m\! (B_{k}\cup\{m\}, \mathcal{B}') \Leftrightarrow u(\mathcal{B})\!<\!u(\mathcal{B}'),
\end{equation}
where the coalition structure is transformed from $\mathcal{B}$ to $\mathcal{B}'$ based on the activation of antenna $m$, i.e., $\mathcal{B'}=\mathcal{B}\backslash B_k\cup\{B_k\cup\{m\}\}$. In the case that pinching antenna $m$ on waveguide $k$ is activated, it can be deactivated based on the following preference relation:
\begin{equation}\label{antennadeact}
(B_k, \mathcal{B}) \!\prec_m\! (B_{k}\backslash\{m\}, \mathcal{B}') \Leftrightarrow u(\mathcal{B})\!<\!u(\mathcal{B}'),
\end{equation}
where $\mathcal{B}'=\mathcal{B}\backslash B_k\cup\{B_k\backslash\{m\}\}$. During the transformation of coalition structures, the channel vectors of all users are changed, and hence, the SIC decoding orders and power allocation for all waveguides need to be re-calculated accordingly. For the antenna activation game, the Nash stability can be defined as follows:
\begin{definition}\label{stability2}
In the antenna activation game, a coalition structure $\mathcal{B}$ is Nash stable if 
\begin{enumerate}
\item $(B_k, \mathcal{B})\succeq_m (B_k\cup\{m\}, \mathcal{B}'), \forall m \in B_k$, or
\item $(B_k, \mathcal{B})\succeq_m (B_k\backslash\{m\}, \mathcal{B}'), \forall m \notin B_k$.
\end{enumerate}\end{definition}

\begin{algorithm}[t]
\caption{Coalitional Game based Waveguide Assignment and Antenna Activation Algorithm}
\label{cgalg}
\begin{algorithmic}[1]
\STATE \textbf{Initialization:}
\STATE Assign all users to the nearest waveguides to obtain $\mathcal{A}$.
\STATE Activate the nearest antenna for each user to obtain $\mathcal{B}$.
\STATE \textbf{Main Loop:}
\FOR{$k\in\mathcal{K}$}
\FOR{$n\in\mathcal{N}$}
\IF{$n\notin A_k$}
\STATE User $n$ performs action following \eqr{userpref}.
\ENDIF 
\ENDFOR
\IF{$|A_k|\neq 0$}
\FOR{$m\in\mathcal{M}$}
\IF{$m\notin B_k$}
\STATE Antenna $m$ performs action following \eqr{antennaact}.
\ELSE
\IF{$|B_k|\neq 1$}
\STATE Antenna $m$ performs action following \eqr{antennadeact}.
\ENDIF 
\ENDIF 
\ENDFOR
\ENDIF
\ENDFOR
\end{algorithmic}
\end{algorithm}

Based on the preference relations of users and pinching antennas, waveguide assignment and antenna activation can be jointly solved by a coalitional game based algorithm, as shown in \textbf{Algorithm \ref{cgalg}}. The proposed algorithm can be started from any coalition structure. In order to reduce the number of required iterations, in the initialization stage, each user is assigned to its nearest waveguide, and its nearest pinching antenna is activated. 

During the main loop, waveguide assignment and antenna activation are iteratively performed for each waveguide. During the waveguide assignment phase, as shown in lines 6 to 10 of \textbf{Algorithm \ref{cgalg}}, any user $n$ not assigned to waveguide $k$ can compare the preference between waveguide $k$ and its currently assigned waveguide based on \eqr{userpref}. If user $n$ prefers waveguide $k$ over its currently assigned waveguide, it can split from its current coalition and merge into coalition $A_k$. For antenna activation, it is implemented on waveguide $k$ if there are users assigned to this waveguide, as shown in line 11. At this point, all pinching antennas on waveguide $k$ can be sequentially activated or deactivated according to \eqr{antennaact} or \eqr{antennadeact}, depending on their previous states. For deactivation, it should be noted that the deactivated antenna cannot be the last activated antenna, as shown in line 16, otherwise this waveguide cannot provide services to the assigned users. The aforementioned steps are is executed repeatedly until neither structure $\mathcal{A}$ nor $\mathcal{B}$ changes in a complete loop. At this stage, the final structures is output, from which the solution of waveguide assignment and antenna activation, i.e., $\boldsymbol{\alpha}$ and $\boldsymbol{\beta}$, can be obtained.
%%%%%%%%%%%%%%%%%%%%%%%%%%%%%%%%%%%%%%%%%%%%%%%%%
\subsection{Properties of Coalitional Game based Algorithm}
In this subsection, the properties of the proposed coalitional game based algorithm are analyzed.

\subsubsection{Complexity}
The complexity of the proposed coalitional game based algorithm can be analyzed by considering the worst case, where all users and pinching antennas need to compare preference. For any waveguide, there are at most $N$ users for waveguide assignment and $M$ antennas for antenna activation, and therefore, for all $K$ waveguides, $KMN$ calculations are performed within the main loop, i.e., a complete cycle. Given the number of cycles $T_c$, the complexity of the waveguide assignment and antenna activation algorithm can be expressed as $\mathcal{O}(T_cKMN)$.

\subsubsection{Convergence}
For waveguide assignment, starting from any initial structure $\mathcal{A}_\mathrm{init}$, the proposed coalitional game based algorithm is able to converge to a final structure $\mathcal{A}_\mathrm{final}$, and the transformation of coalition structures can be expressed as follows:
\begin{equation}
\mathcal{A}_\mathrm{init}\to\mathcal{A}_1\to\mathcal{A}_2\to\cdots\to\mathcal{A}_\mathrm{final}.
\end{equation}
Assuming $\mathcal{A}_a$ and $\mathcal{A}_b$ are two adjacent coalition structures, where $b=a+1$, then, from $\mathcal{A}_a$ to $\mathcal{A}_b$, there exists one user who splits from a coalition and merges into another coalition. This operation follows \eqr{userpref}, which implies that the sum rate is strictly increased, i.e., $v(\mathcal{A}_a)<v(\mathcal{A}_b)$. For antenna activation following \eqr{antennaact} or \eqr{antennadeact}, a similar conclusion can be obtained, that is, the activation or deactivation of a pinching antenna can strictly increase the sum rate. Since both the number of users and the number of pinching antennas are limited, the number of possible structures is equal to the Bell number \cite{ray2007game}, and the upper bound on the sum rate are finite. Therefore, with the strict increase of the sum rate, the coalition structures of both waveguide assignment and antenna activation are guaranteed to converge to a final structure whose coalition value cannot be further increased.

\subsubsection{Stability}
Based on Definition \ref{stability1} and Definition \ref{stability2}, the obtained final structure is not Nash stable if there is a user or pinching antenna in a coalition prefers a new coalition structure. Based on the preference relation \eqr{userpref}, \eqr{antennaact}, or \eqr{antennadeact}, this operation can strictly increase the coalition value. However, this contradicts the conclusion that the coalition value of the final structure cannot be further increased. Therefore, the final structure obtained from the proposed coalitional game based algorithm is always Nash stable.

%%%%%%%%%%%%%%%%%%%%%%%%%%%%%%%%%%%%%%%%%%%%%%%%%
%%%%%%%%%%%%%%%%%%%%%%%%%%%%%%%%%%%%%%%%%%%%%%%%%
\section{Solutions for Power Allocation}
In this section, the power allocation problem in \eqr{paproblem} is studied with the given solution of waveguide assignment, antenna activation and SIC decoding order design, where monotonic optimization and SCA are employed to provide solutions with different performance-complexity trade-offs, respectively.

%%%%%%%%%%%%%%%%%%%%%%%%%%%%%%%%%%%%%%%%%%%%%%%%%
\subsection{Monotonic Optimization based Power Allocation}\label{mosolution}
Based on the definition of increasing functions \cite{zhang2013monotonic}, it can be obtained that in the considered NOMA-based pinching-antenna system, the sum rate is a monotonically increasing function of the power allocation coefficients. Therefore, problem \eqr{paproblem} can be regarded as a monotonic maximization problem, and monotonic optimization is utilized to find the global optimal solution. 

\begin{figure}[t]
\centering{
\subfigure[]{\centering{\includegraphics[width=40mm]{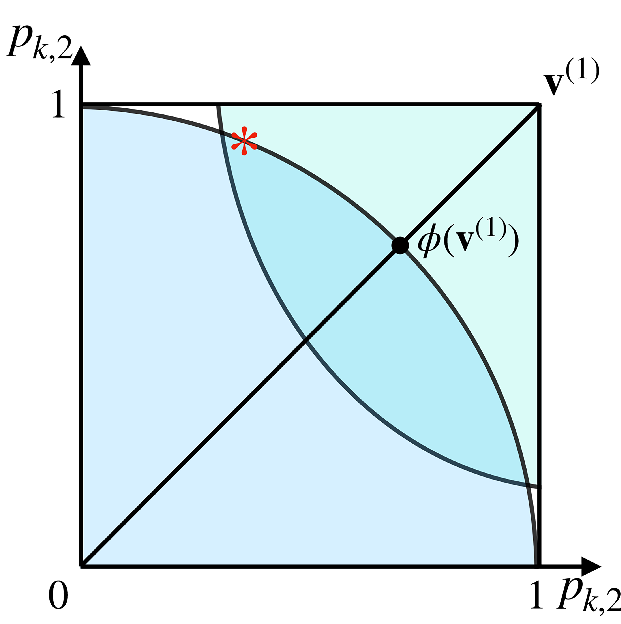}}}
\subfigure[]{\centering{\includegraphics[width=40mm]{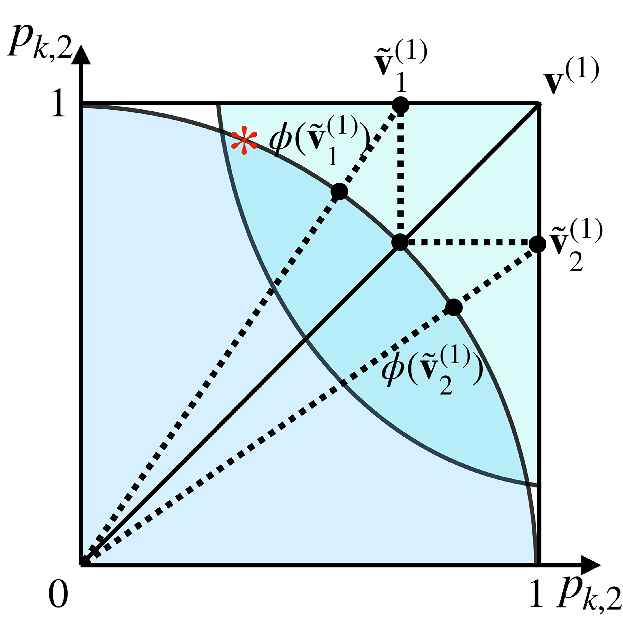}}}
\subfigure[]{\centering{\includegraphics[width=40mm]{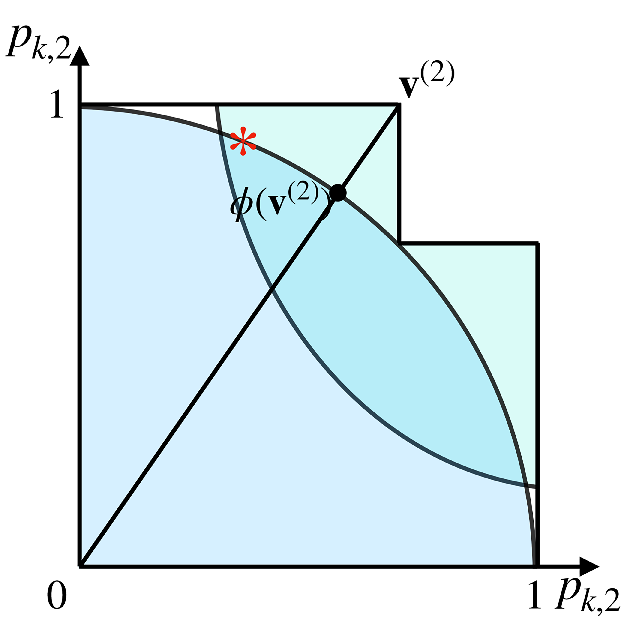}}}
\subfigure[]{\centering{\includegraphics[width=40mm]{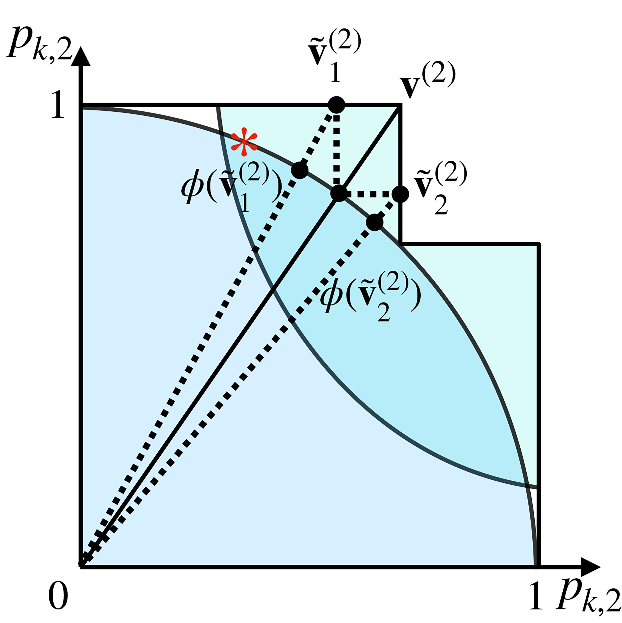}}}}
\caption{An illustration of Algorithm \ref{alg2} with two users. The blue area is the feasible set $\mathcal{G}$, the green area is the feasible set $\mathcal{H}$, and the red star is the optimal point.}
\label{mo}
\end{figure}

Before transforming the problem into the canonical monotonic optimization formulation, problem \eqr{paproblem} can be further decoupled to reduce the complexity. Due to the fact that the sum rate is a monotonically increasing function of the power allocation coefficients, in order to maximize the sum rate, all available transmit power needs to be exploited if there are assigned users on the waveguide, i.e., $\sum_{n=1}^N\alpha_{k,n}p_{k,n} \!=\! 1$. On the other hand, for the waveguides without any users to be served, $\sum_{n=1}^N\alpha_{k,n}p_{k,n} \!=\! 0$ is satisfied. As a result, with the given strategy of waveguide assignment, the sum of power allocation coefficients is fixed, and therefore, the power allocation coefficients can be removed from the inter-waveguide interference, which implies that problem \eqr{paproblem} can be decoupled into a series of independent sub-problems. For the users assigned to waveguide $k$, i.e., $A_k$, where $|A_k|>0$, the power allocation problem can be presented as follows:
\begin{subequations}
\begin{empheq}{align}
\max_{\mathbf{p}_k}\quad & \sum_{n\in A_k}R_{k,n}^{(\mathrm{Q}_k)}\\
\textrm{s.t.} \quad & R_{k,n}^{(\mathrm{Q}_k)}\ge R_\mathrm{min}, \forall n\in A_k,\\
& 0\le p_{k,n}, \forall n\in A_k,\\ 
& \sum\nolimits_{n\in A_k}p_{k,n} \le 1,
\end{empheq}
\label{paproblem1}
\end{subequations}\vspace{-2mm}\\
where $\mathbf{p}_k$ is the set of power allocation coefficients of the users assigned to waveguide $k$. According to constraints (\ref{paproblem1}c) and (\ref{paproblem1}d), it can be obtained that $\mathbf{p}_k\in[\mathbf{0},\mathbf{1}]$. Moreover, constraints (\ref{paproblem1}b) and (\ref{paproblem1}d) can be respectively rewritten as $h_n(\mathbf{p}_k)\ge 0, \forall n\in A_k$ and $g(\mathbf{p}_k)\le 0$, where 
\begin{equation}
h_n(\mathbf{p}_k)\triangleq R_{k,n}^{(\mathrm{Q}_k)}-R_{\min},
\end{equation}
and
\begin{equation}
g(\mathbf{x})\triangleq\sum_{n\in\mathcal{A}_k}p_{k,n}-1.
\end{equation}
Therefore, the power allocation problem in \eqr{paproblem1} can be represented as a canonical monotonic optimization form, as shown below:
\begin{subequations}
\begin{empheq}{align}
\max_{\mathbf{p}_k}\quad\!\! & f(\mathbf{p}_k)\\
\textrm{s.t.} \quad & \mathbf{p}_k\in \mathcal{G}\cap\mathcal{H},\notag
\end{empheq}
\label{canonical}
\end{subequations}\vspace{-2mm}\\
where
\begin{equation}
f(\mathbf{p}_k)=\!\sum_{n\in A_k}R_{k,n}^{(\mathrm{Q}_k)},
\end{equation}
\begin{equation}\label{defg}
\mathcal{G} = \left\{\mathbf{p}_k|\mathbf{p}_k\in[\mathbf{0}, \mathbf{1}], g_k(\mathbf{p}_k)\le 0\right\},
\end{equation}
and
\begin{equation}
\mathcal{H} = \left\{\mathbf{p}_k|\mathbf{p}_k\in[\mathbf{0}, \mathbf{1}], h_n(\mathbf{p}_k)\ge 0\right\}.
\end{equation}
In problem \eqr{canonical}, $\mathcal{G}$ and $\mathcal{H}$ respectively represent the normal set and conormal set \cite{sun2017noma}, and the feasible set is the intersection of these two sets, as shown in \fref{mo}(a). Furthermore, because $f(\mathbf{p}_k)$ is an increasing function of $\mathbf{p}_k$, the global optimum is located on the boundary of the feasible set, which is, however, unknown. In this section, by constructing a series of polyblocks, a polyblock outer approximation algorithm can be used to approach the boundary, as shown in \textbf{Algorithm \ref{alg2}}.

\begin{algorithm}[t]
\caption{Polyblock Outer Approximation Algorithm}
\label{alg2}
\begin{algorithmic}[1]
\STATE Initialize vertex set $\mathcal{V}^{(1)}=\{\mathbf{v}^{(1)}\}$ and polyblock $\mathcal{P}^{(1)}$.
\STATE Initialize $R_\mathrm{max}=-\infty$, $\theta=1$, and $\epsilon$.
\IF{$|f(\boldsymbol{\phi}(\mathbf{v}^{(\theta)}))-R_\mathrm{max}|>\epsilon$}
\STATE Obtain projection $\boldsymbol{\phi}(\mathbf{v}^{(\theta)})$ based on \eqr{projection}.
\IF{$\boldsymbol{\phi}(\mathbf{v}^{(\theta)})\in\mathcal{G}\cap\mathcal{H}$ and $f(\boldsymbol{\phi}(\mathbf{v}^{(\theta)}))\ge R_\mathrm{max}$}
\STATE Set $\mathbf{p}_k^*=\boldsymbol{\phi}(\mathbf{v}^{(\theta)})$ and $R_\mathrm{max}=f(\boldsymbol{\phi}(\mathbf{v}^{(\theta)}))$.
\ENDIF
\STATE Calculate vertex set $\tilde{\mathcal{V}}^{(\theta)}=\{\tilde{\mathbf{v}}_i^{(\theta)}| \forall i\}$ from \eqr{vertexupdate}.
\STATE Update vertex set $\mathcal{V}^{(\theta+1)}$ based on \eqr{vertexsetupdate}.
\STATE Construct polyblock $\mathcal{P}^{(\theta+1)}$ with vertex set $\mathcal{V}^{(\theta+1)}$.
\STATE Find vertex $\mathbf{v}^{(\theta+1)}$ from $\mathcal{V}^{(\theta+1)}$ based on \eqr{nextvertex}.
\STATE Set $\theta=\theta+1$.
\ENDIF
\STATE Output $\mathbf{p}_k^*$.
\end{algorithmic}
\end{algorithm}

In the initialization phase of \textbf{Algorithm \ref{alg2}}, an initial polyblock $\mathcal{P}^{(1)}$ is constructed from vertex set $\mathcal{V}^{(1)}$ containing vertex $\mathbf{v}^{(1)}$, where all elements are set to $1$, as shown in \fref{mo}(a). Based on vertex $\mathbf{v}^{(1)}$, a projection on the upper boundary of the feasible set can be obtained and denoted by $\boldsymbol{\phi}(\mathbf{v}^{(1)})$. In order to calculate the upper boundary point, a ratio coefficient $\zeta \in (0,1)$ is defined and the projection of vertex $\mathbf{v}^{(1)}$ is given by $\boldsymbol{\phi}(\mathbf{v}^{(1)})=\zeta\mathbf{v}^{(1)}$. At any iteration $\theta$, $\zeta$ can be obtained by solving the following problem:
\begin{equation}
\zeta = \max\{\tilde{\zeta}\in(0,1)|\tilde{\zeta}\mathbf{v}^{(\theta)}\in\mathcal{G}\}.
\end{equation}
From \eqr{defg}, the above function can be transformed as follows:
\begin{equation}
\zeta = \max\{\tilde{\zeta}\in(0,1)|g(\tilde{\zeta}\mathbf{v}^{(\theta)})\le 0\}
\end{equation}
It can be observed that $g(\tilde{\zeta}\mathbf{v}^{(\theta)})$ is monotonically increasing with $\tilde{\zeta}$, $\tilde{\zeta}\sum_{n\in A_k} p_{k,n} = 1$ always holds at the upper boundary point, which implies the following equation:
\begin{equation}\label{projection}
\zeta = \frac{1}{\sum_{n\in A_k} p_{k,n}}.
\end{equation}
At this stage, a new vertex set $\mathcal{\tilde{V}}^{(\theta)}=\{\mathbf{\tilde{v}}_1^{(\theta)}, \mathbf{\tilde{v}}_2^{(\theta)}, \cdots, \mathbf{\tilde{v}}_N^{(\theta)}\}$ can be obtained based on the following equation:
\begin{equation}\label{vertexupdate}
\tilde{\mathbf{v}}_i^{(\theta)}=\mathbf{v}^{(\theta)}-(v_i^{(\theta)}-\phi_i(\mathbf{v}^{(\theta)}))\mathbf{e}_i,
\end{equation}
where $v_i^{(\theta)}$ is the $i$-th element of $\mathbf{v}^{(\theta)}$, $\phi_i(\mathbf{v}^{(\theta)})$ is the $i$-th element of $\boldsymbol{\phi}(\mathbf{v}^{(\theta)})$, and $\mathbf{e}_i$ is the $i$-th unit vector. As shown in \fref{mo}(b), in the case that there are two users assigned to waveguide $k$, two new vertices $\tilde{\mathbf{v}}_1^{(1)}$ and $\tilde{\mathbf{v}}_2^{(1)}$ are obtained to replace the original vertex $\mathbf{v}_1^{(1)}$, and a new polyblock $\mathcal{P}^{(2)}$ can be constructed, as shown in \fref{mo}(c). In order to improve the convergence rate by reducing the size of the vertex set, at any iteration $\theta$, all suboptimal vertices can be removed, and the vertex set is updated as follows:
\begin{equation}\label{vertexsetupdate}
\mathcal{V}^{(\theta+1)}=\{\mathcal{V}^{(\theta)}\backslash\hat{\mathcal{V}}^{(\theta)}\}\cup\tilde{\mathcal{V}}^{(\theta)},
\end{equation}
where $\hat{\mathcal{V}}^{(\theta)}=\{\mathbf{v}\in\mathcal{V}^{(\theta)}|\mathbf{v}>\phi_i(\mathbf{v}^{(\theta)})\}$. After that, the next vertex $\mathbf{v}^{(\theta+1)}$ is selected from $\mathcal{V}^{(\theta+1)}$, whose projection can achieve the maximum sum rate, i.e.,
\begin{equation}\label{nextvertex}
\mathbf{v}^{(\theta+1)}=\argmax\{f(\boldsymbol{\phi}(\mathbf{v}))|\mathbf{v}\in\mathcal{V}^{(\theta+1)}\}.
\end{equation}
Following this procedure, a sequence of polyblocks can be constructed to approximate the feasible set, as follows:
\begin{equation}
\mathcal{P}^{(1)}\supset\mathcal{P}^{(2)}\supset\mathcal{P}^{(3)}\supset\cdots\supset \mathcal{G}\cap\mathcal{H}.
\end{equation}

The complexity of \textbf{Algorithm \ref{alg2}} is mainly determined by line 11 and the number of iterations \cite{kaidi2025fl}. In particular, at the $\theta$-th iteration, by calculating the projection $\boldsymbol{\phi}(\mathbf{v}^{(\theta)})$, $N_k$ new vertices can be obtained and added to the vertex set $\mathcal{V}^{(\theta+1)}$. Hence, in the worst case that no vertices can be removed, the size of the vertex set at the $\theta+1$-th iteration is $\theta N_k$. With a given number of iterations $T_p$, $\sum_{\theta=1}^{T_p-1}\theta N_k$ calculations are performed by \textbf{Algorithm \ref{alg2}}, and therefore, the complexity can be expressed as $\mathcal{O}(T_p^2N_k)$.

According to \cite{tuy2000monotonic, zhang2013monotonic}, the polyblock outer approximation algorithm is able to converge to the global optimum within sufficient iterations. However, when the number of assigned users in waveguide $k$ is large, the number of required iterations exponentially increases with the vertex set, i.e., $\mathcal{V}^{(\theta)}$. Therefore, the monotonic optimization based algorithm is included to find the upper bound on the sum rate of the considered system, and a low-complexity algorithm is proposed in the next subsection.

%%%%%%%%%%%%%%%%%%%%%%%%%%%%%%%%%%%%%%%%%%%%%%%%%
\subsection{SCA based Sub-Optimal Solution}
Due to the high complexity of the monotonic optimization based algorithm, a SCA based algorithm is proposed in this subsection to provide the low-complexity sub-optimal solution. Recall that in problem \eqr{paproblem1}, the users assigned to each waveguide and their decoding order are determined. Assuming there are $N_k$ users in cluster $A_k$ and the decoding order $Q_k$ is from $1$ to $N_t$, based on the proof in Appendix~A, the data rate of the $n$-th decoded user\footnote{Note that in this subsection, all variables are defined in terms of the SIC decoding order in $Q_k$, rather than the indices of users.} assigned to waveguide $k$ can be expressed below:
\begin{equation}
R_{k,n}^{(\mathrm{Q}_k)}\!=\!\log_2\!\left(\!1\!+\!\frac{p_{k,n}\frac{P_t}{M_k}}{\frac{P_t}{M_k}\!\sum_{i=n+1}^{N_k}\!p_{k,i}\!+\!C_{k,n}}\!\right),
\end{equation}
where $C_{k,n}^{(\mathrm{Q}_k)}$ is a constant as follows:
\begin{equation}
C_{k,n}^{(\mathrm{Q}_k)}=\max\left\{\frac{I_{k,n'}\!+\sigma^2}{|h_{k,n'}|^2}\Big|\forall n'\ge n\right\}.
\end{equation}
Moreover, the last decoded user can decode its signal with the following data rate:
\begin{equation}
R_{k,N_k}^{(\mathrm{Q}_k)}\!=\!\log_2\!\left(\!1\!+\!\frac{p_{k,N_k}\frac{P_t}{M_k}}{C_{k,N_k}}\!\right).
\end{equation}
Therefore, problem \eqr{paproblem1} can be rewritten as follows:
\begin{subequations}
\begin{empheq}{align}
\max_{\mathbf{p}_k}\quad\!\!\!\! & \sum_{n\in A_k}\!R_{k,n}^{(\mathrm{Q}_k)}\\
\textrm{s.t.} \quad & \frac{p_{k,n}\frac{P_t}{M_k}}{\frac{P_t}{M_k}\!\!\sum_{i=n+1}^{N_k}\!p_{k,i}\!+\!C_{k,n}^{(\mathrm{Q}_k)}}\!\ge\!\gamma_\mathrm{min}, \forall n \!\in\! A_k, n \!\neq\! N_k,\!\!\!\\
& p_{k,N_k}\frac{P_t}{M_k}\!\ge\!\gamma_\mathrm{min}C_{k,N_k}^{(\mathrm{Q}_k)},\!\!\!\\\nonumber
& \text{(\ref{paproblem1}c), and (\ref{paproblem1}d)},
\end{empheq}
\label{paproblem2}
\end{subequations}\vspace{-2mm}\\
where $\gamma_\mathrm{min} = (2^{R_\mathrm{min}}\!-\!1)$. Since the objective function (\ref{paproblem2}a) is non-convex, a set of slack variables $\boldsymbol{\gamma}_k$ is introduced, and problem \eqr{paproblem2} can be equivalently transformed as follows:
\begin{subequations}
\begin{empheq}{align}
\max_{\mathbf{p}_k,\boldsymbol{\gamma}_k}\quad\!\!\!\! & \sum_{n\in A_k}\log_2(1+\gamma_{k,n})\\
\textrm{s.t.} \quad & \frac{p_{k,n}\frac{P_t}{M_k}}{\frac{P_t}{M_k}\!\!\sum_{i=n+1}^{N_k}\!p_{k,i}\!+\!C_{k,n}^{(\mathrm{Q}_k)}}\!\ge\! \gamma_{k,n}, \forall n\!\in\! A_k, n \!\neq\! N_k,\!\!\!\\
& p_{k,N_k}\frac{P_t}{M_k}\!\ge\! \gamma_{k,N_k}C_{k,N_k}^{(\mathrm{Q}_k)},\\\nonumber
& \text{(\ref{paproblem1}c), (\ref{paproblem1}d), (\ref{paproblem2}b), and (\ref{paproblem2}c)}.
\end{empheq}
\label{paproblem3}
\end{subequations}\vspace{-2mm}\\
In the above problem, constraints (\ref{paproblem2}b) and (\ref{paproblem2}c) correspond to the users suffering from intra-waveguide interference and the user able to remove intra-waveguide interference, respectively. By introducing another set of slack variables $\boldsymbol{\mu}_k$, the following problem can be obtained:
\begin{subequations}
\begin{empheq}{align}
\max_{\mathbf{p}_k,\boldsymbol{\gamma}_k,\boldsymbol{\mu}_k}\quad\!\!\!\!\! & \sum_{n\in A_k}\!\log_2(1+\gamma_{k,n})\\
\textrm{s.t.} \quad & \mu_{k,n}\!\ge\!\frac{P_t}{M_k}\!\!\sum_{i=n+1}^{N_k}\!\!\!p_{k,i}\!+\!C_{k,n}^{(\mathrm{Q}_k)},\forall n\!\in\! A_k, n \!\neq\! N_k,\!\!\!\\
& p_{k,n}\!\frac{P_t}{M_k}\!\ge\!\mu_{k,n}\gamma_{k,n}, \forall n\!\in\! A_k, n \!\neq\! N_k,\\\nonumber
& \text{(\ref{paproblem1}c), (\ref{paproblem1}d), (\ref{paproblem2}b), (\ref{paproblem2}c), and (\ref{paproblem3}c)}.
\end{empheq}
\label{paproblem4}
\end{subequations}\vspace{-2mm}\\
Note that this problem is still non-convex due to the terms $\mu_{k,n}\gamma_{k,n}$ in constrain (\ref{paproblem4}c), and the first order Taylor expansion is employed for approximation, as follows:
\begin{equation}
\mu_{k,n}\gamma_{k,n}\ge\mu_{k,n}^{(t)}\gamma_{k,n}^{(t)}\!+\!\gamma_{k,n}^{(t)}(\mu_{k,n}\!-\!\mu_{k,n}^{(t)})\!+\!\mu_{k,n}^{(t)}(\gamma_{k,n}\!-\!\gamma_{k,n}^{(t)}),
\end{equation}
where $\mu_{k,n}^{(t)}$ and $\gamma_{k,n}^{(t)}$ are the values of $\mu_{k,n}$ and $\gamma_{k,n}$ at the $t$-th iteration. In this case, constraint (\ref{paproblem4}c) can be approximated by utilizing the initial values $\mu_{k,n}^{(0)}$ and $\gamma_{k,n}^{(0)}$. As a result, problem \eqr{paproblem4} can be recast as follows:
\begin{subequations}
\begin{empheq}{align}
\max_{\mathbf{p}_k,\boldsymbol{\gamma}_k,\boldsymbol{\mu}_k}\quad\!\!\!\!\! & \sum_{n\in A_k}\!\log_2(1+\gamma_{k,n})\\\nonumber
\textrm{s.t.} \quad & p_{k,n}\!\frac{P_t}{M_k}\!\ge\!\mu_{k,n}^{(t)}\gamma_{k,n}^{(t)}\!+\!\gamma_{k,n}^{(t)}(\mu_{k,n}\!-\!\mu_{k,n}^{(t)})\\
&\quad\quad+\mu_{k,n}^{(t)}(\gamma_{k,n}\!-\!\gamma_{k,n}^{(t)}),\forall n\!\in\! A_k, n \!\neq\! N_k,\\\nonumber
& \text{(\ref{paproblem1}c), (\ref{paproblem1}d), (\ref{paproblem2}b), (\ref{paproblem2}c), (\ref{paproblem3}c), and (\ref{paproblem4}b)}.
\end{empheq}
\label{paproblem5}
\end{subequations}\vspace{-2mm}\\

This problem is convex and can therefore be directly solved by convex solvers such as CVX \cite{cvx}. With the given initial power allocation coefficients, an SCA based algorithm is proposed in \textbf{Algorithm \ref{alg3}}. It is worth noting that the performance of SCA is highly dependent on the initial power allocation coefficients and cannot be guaranteed to converge to the global optimum. However, for the same number of optimization variables and the same value of $\epsilon$, the number of iterations required by SCA is significantly smaller than that of the polyblock outer approximation algorithm. Thus, it serves as a low-complexity alternative to the algorithm based on monotonic optimization.

\begin{algorithm}[t]
\caption{SCA based algorithm for solving problem \eqr{paproblem5}}
\label{alg3}
\begin{algorithmic}[1]
\STATE Initialize $\boldsymbol{\mu}_k^{(0)}$ and $\boldsymbol{\gamma}_k^{(0)}$ with the fixed power allocation.
\STATE Initialize $R_\mathrm{opt}=-\infty$, $t=0$, and $\epsilon$.
\IF{$|\sum_{n\in A_k}\!\log_2(1\!+\!\gamma_{k,n}^{(t)})-R_\mathrm{opt}|>\epsilon$}
\STATE Obtain $\boldsymbol{\mu}_k$, $\boldsymbol{\gamma}_k$, and $\mathbf{p}_k$ by solving problem \eqr{paproblem5}.
\STATE Set $\boldsymbol{\mu}_k^{(t+1)}=\boldsymbol{\mu}_k$ and $\boldsymbol{\gamma}_k^{(t+1)}=\boldsymbol{\gamma}_k$.
\STATE Set $R_\mathrm{opt}=\sum_{n\in A_k}\!\log_2(1\!+\!\gamma_{k,n}^{(t)})$.
\STATE Set $t=t+1$.
\ENDIF
\STATE Set $\mathbf{p}_k^*=\mathbf{p}_k$.
\end{algorithmic}
\end{algorithm}

%%%%%%%%%%%%%%%%%%%%%%%%%%%%%%%%%%%%%%%%%%%%%%%%%
%%%%%%%%%%%%%%%%%%%%%%%%%%%%%%%%%%%%%%%%%%%%%%%%%
\section{Simulation Results}
In this section, simulation results are presented to evaluate the performance of the proposed NOMA-based multi-waveguide pinching-antenna system and the developed optimization solutions. In the simulation setup, the height of each waveguide is $d=3$~m, the length of the rectangular area is $D_x=10$~m, the carrier frequency is $f=28$~GHz, the noise power is $\sigma^2=-90$~dBm, and $n_\mathrm{eff}=1.4$. Based on the number of available pinching antennas, the pinching antennas are evenly distributed along each waveguide, where the antenna spacing is greater than or equal to half of the wavelength.

\begin{figure}[!t]
\centering{\includegraphics[width=84mm]{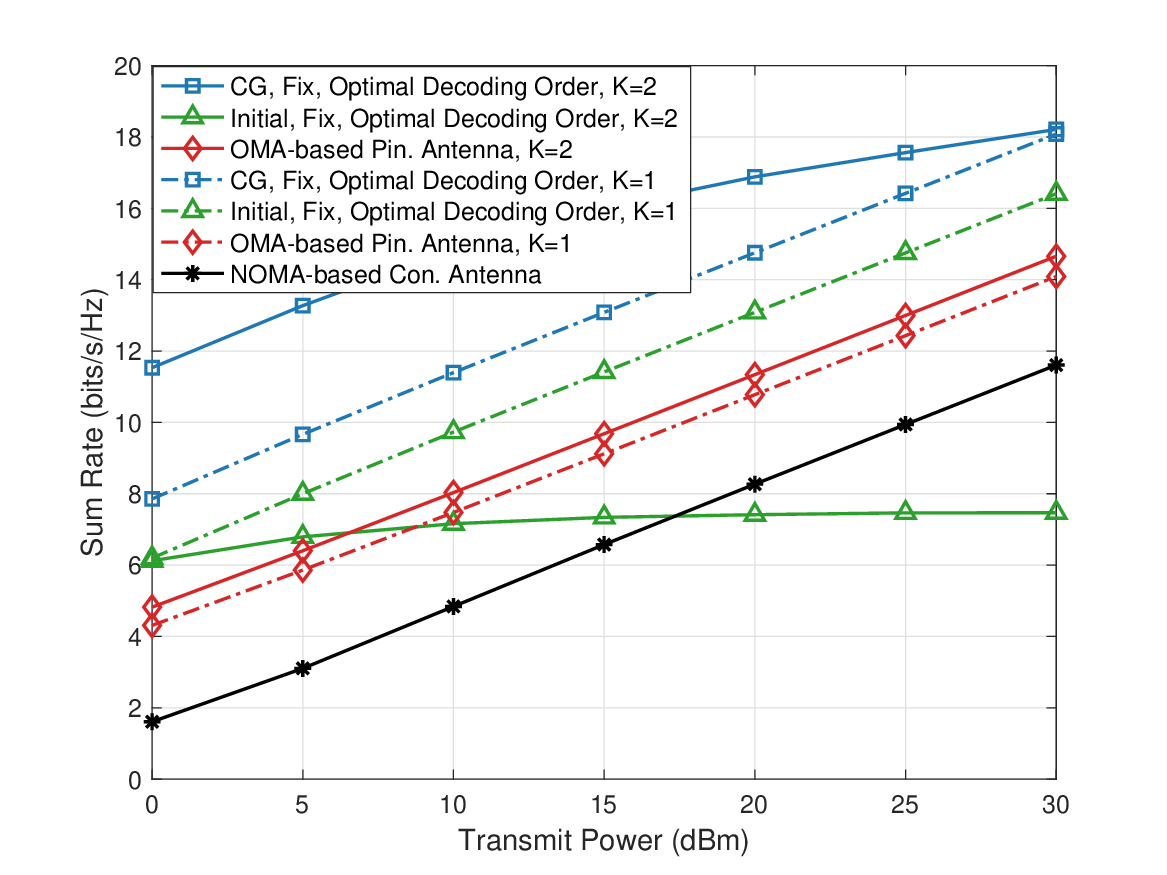}}
\caption{Achievable sum rate with different systems, where $D_y=10$~m, $M=20$, and $N=10$.}
\label{result1}
\end{figure}

The difference between conventional fixed-location antenna systems and pinching-antenna systems is shown in \fref{result1}. In conventional fixed-location antenna systems, $M$ antennas are positioned in the centre with a height of $d$ and an antenna spacing of $\lambda/2$. For pinching-antenna systems, the waveguides are located at $y_k^\mathrm{Pin}=0$ and $y_k^\mathrm{Pin}=\pm 2.5$ for the cases of $K=1$ and $K=2$, respectively. In the OMA scheme, each user is served by the nearest pinching antenna on the nearest waveguide in different time slots. It can be observed from \fref{result1} that the pinching-antenna systems exhibit significant superiority over conventional fixed-location antenna systems, and the proposed coalitional based algorithm (denoted by CG in the figure) further enhances the sum rate. Moreover, the implementation of additional waveguides facilitates the improvement of frequency efficiency, while the overall performance becomes more sensitive to optimization. Specifically, under fixed system parameters, the sum rate of the multi-waveguide system is clearly lower than that of the single-waveguide system, particularly when the transmit power is high. However, based on the coalitional game based algorithm, the performance of the multi-waveguide system can surpass that of the single-waveguide system. Furthermore, \fref{result1} also indicates that the NOMA-based pinching-antenna system can significantly outperform its OMA-based counterpart.

\begin{figure}[!t]
\centering{
\subfigure[Sum Rate]{\centering{\includegraphics[width=84mm]{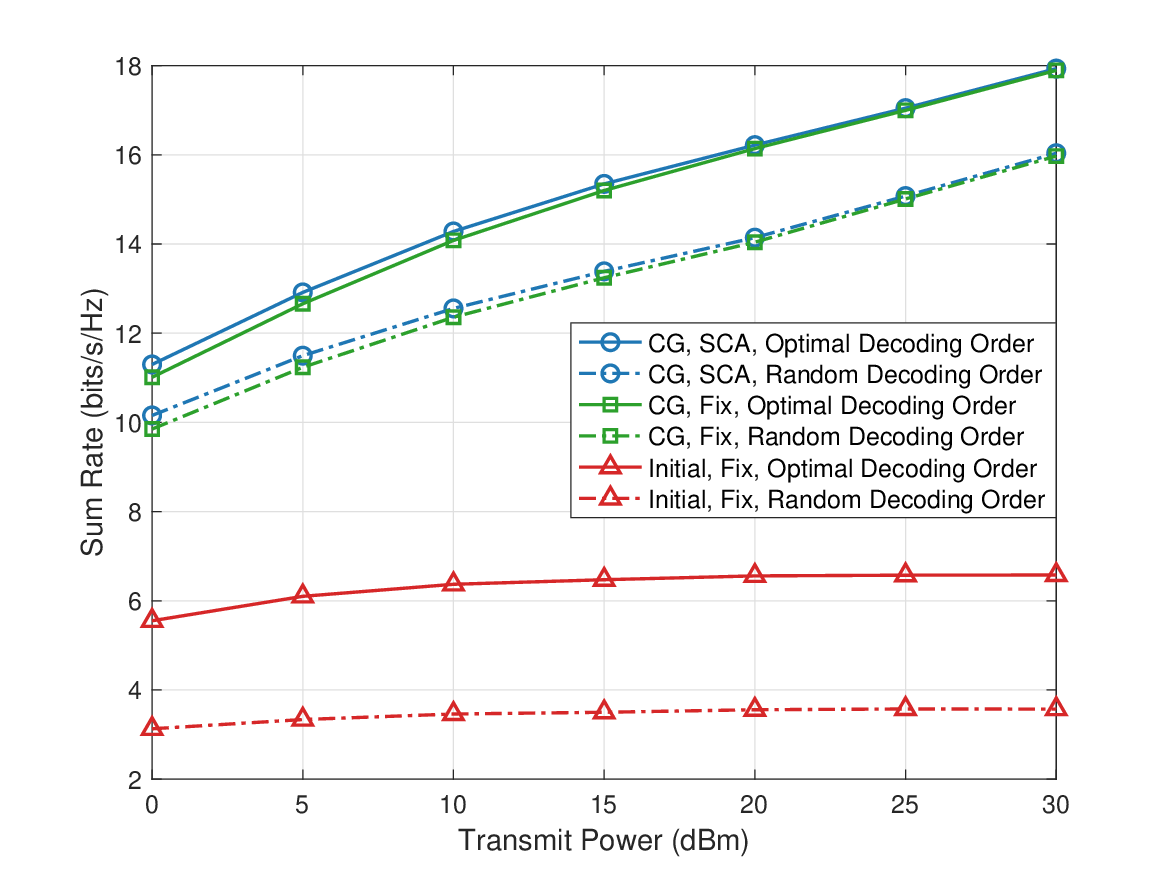}}}
\subfigure[Outage Probability]{\centering{\includegraphics[width=84mm]{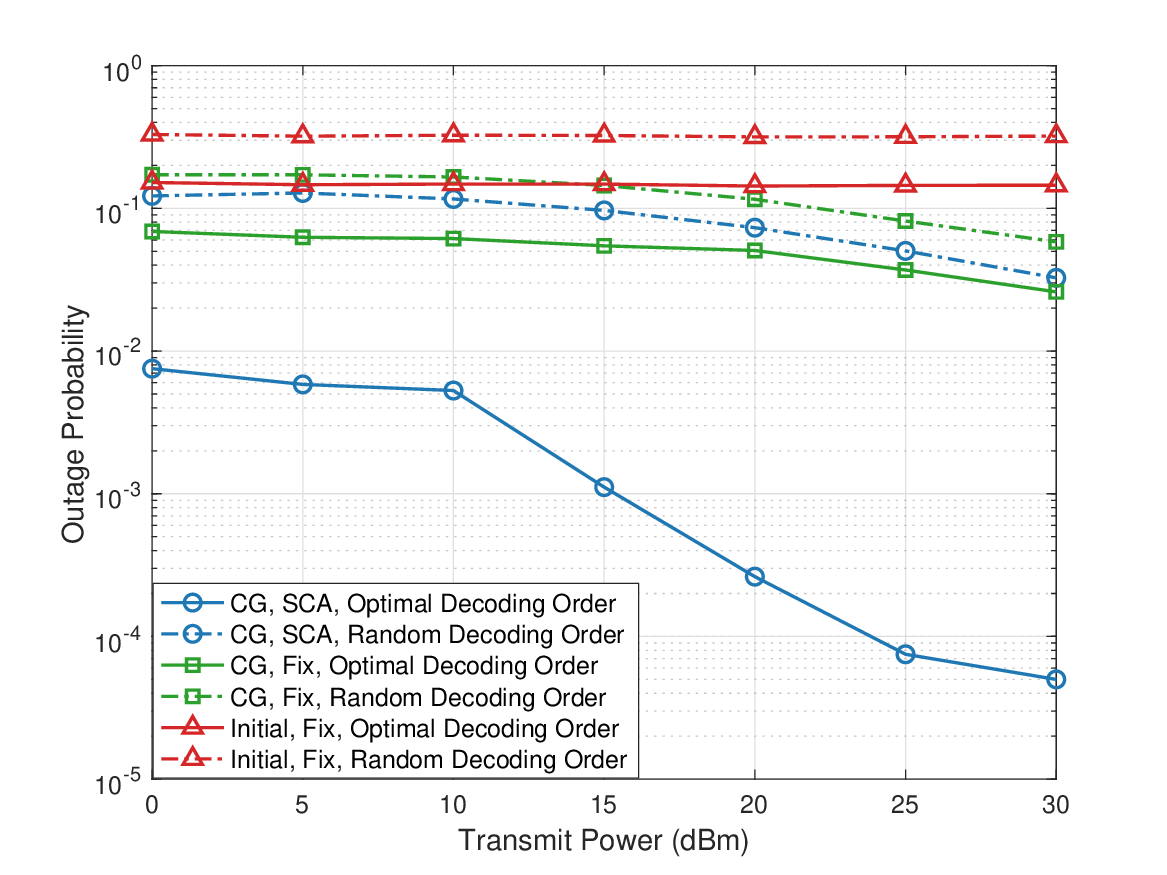}}}}
\caption{Impact of the transmit power on the sum rate and outage probability, $D_y=8$~m, $K=2$, $M=20$, $N=8$, and $R_\mathrm{min} = 0.1$~bps/Hz.}
\label{result2}
\end{figure}

In \fref{result2}, the performance of the proposed optimization solutions is demonstrated. It shows that the coalitional game based waveguide assignment and antenna activation algorithm is able to significantly enhance the sum rate. In contrast, the improvement brought by the SCA based power allocation algorithm is mainly observed in terms of outage probability reduction. In particular, when the transmit power exceeds $20$~dBm, the improvement in the sum rate achieved through the SCA based power allocation algorithm becomes limited, as more power is allocated to the users with poor channel conditions in order to satisfy the QoS requirements. Moreover, the impact of the derived optimal SIC decoding order is illustrated in \fref{result2}. By employing the optimal SIC decoding order, the degradation in achievable data rate caused by SIC decoding can be avoided, resulting in improvements in both the sum rate and outage probability.

\begin{figure}[!t]
\centering{
\subfigure[Sum Rate]{\centering{\includegraphics[width=84mm]{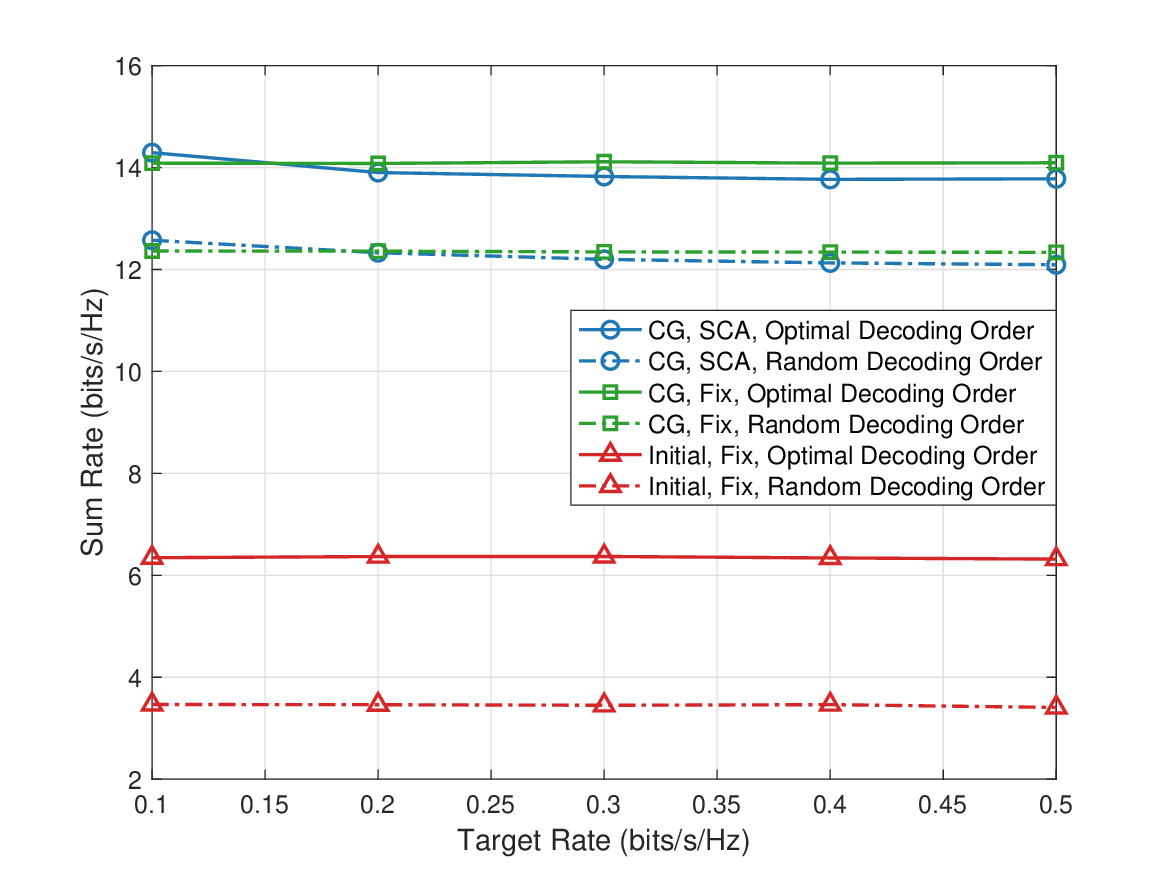}}}
\subfigure[Outage Probability]{\centering{\includegraphics[width=84mm]{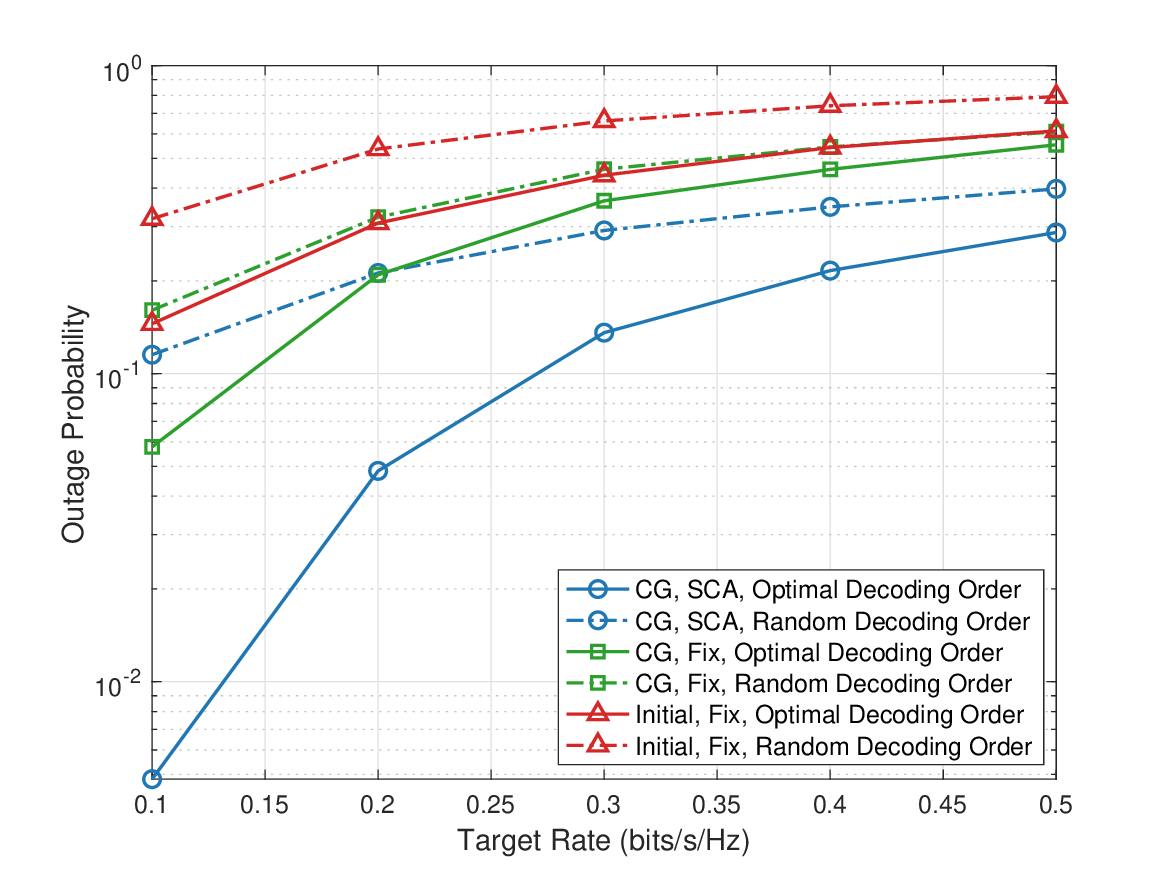}}}}
\caption{Impact of the target rate on the sum rate and outage probability, where $D_y=8$~m, $K=2$, $M=20$, $N=8$, and $P_t=10$~dBm.}
\label{result3}
\end{figure}

\fref{result3} illustrates the impact of the target rate on the performance of the proposed multi-waveguide pinching-antenna system. As the target rate increases, the outage probability of all schemes rises accordingly. Since the waveguide assignment and antenna activation procedures do not incorporate QoS constraints, the sum rate achieved by the coalitional game based algorithm remains unaffected by a change in the target rate. In contrast, the SCA based power allocation algorithm prioritizes satisfying QoS requirements, resulting in a slight reduction in sum rate when the target rate is greater than $0.1$~bits/s/Hz. Additionally, \fref{result3} also demonstrates the effectiveness of the optimal SIC decoding order in enhancing system performance.

\begin{figure}[!t]
\centering{
\subfigure[Sum Rate]{\centering{\includegraphics[width=84mm]{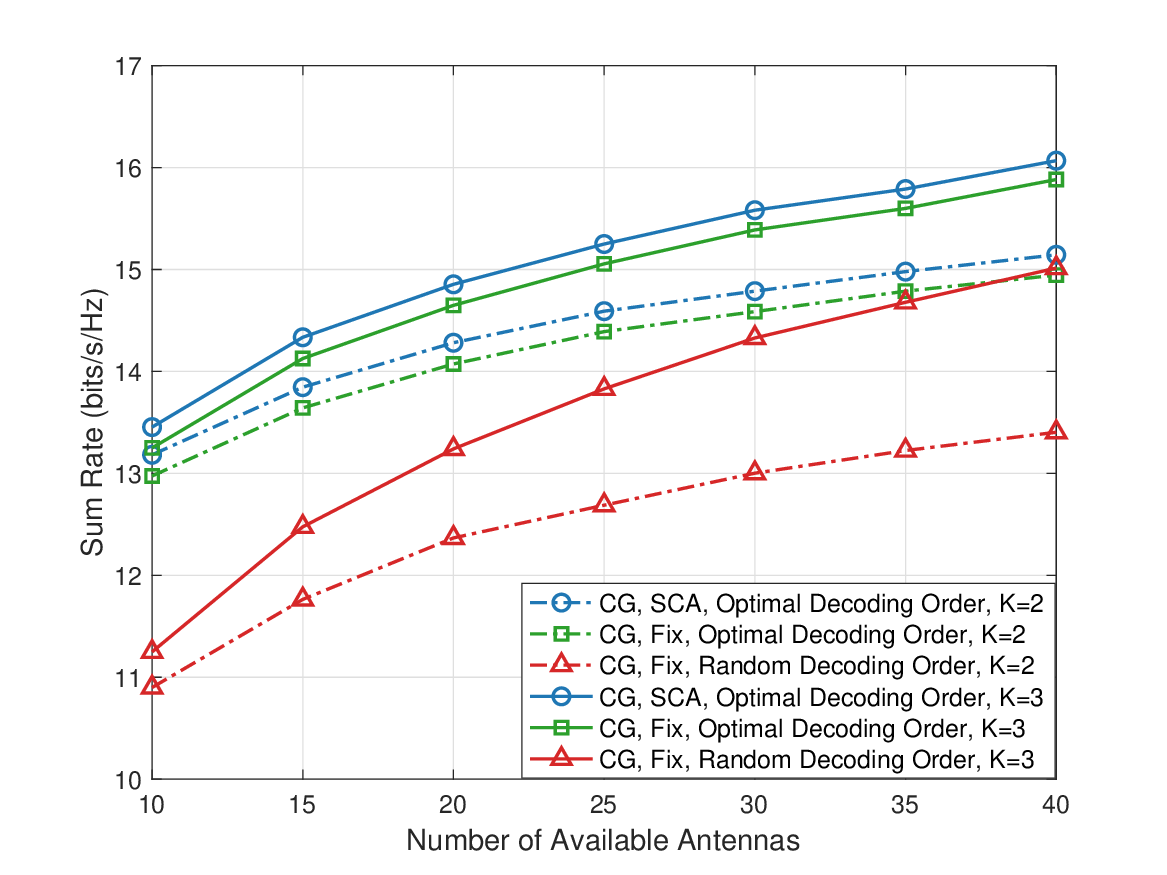}}}
\subfigure[Number of Activated Antennas]{\centering{\includegraphics[width=84mm]{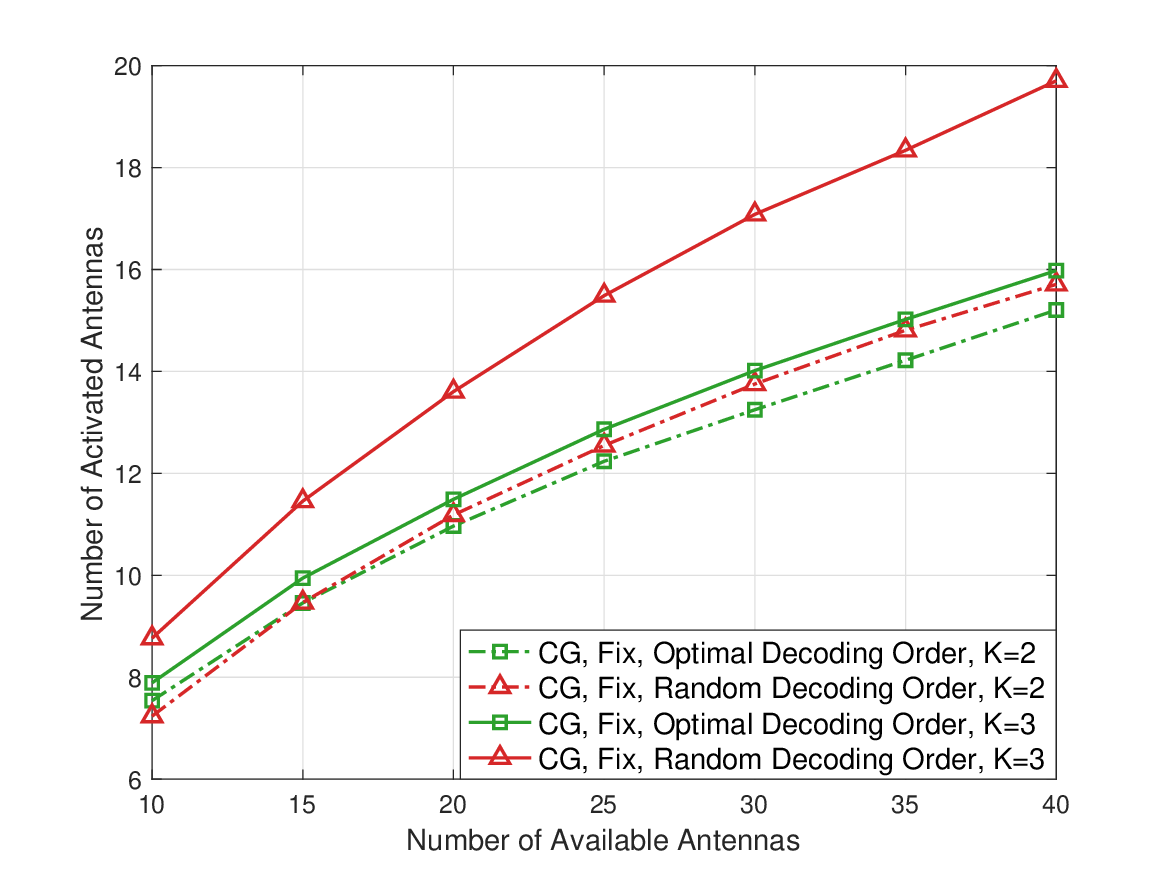}}}}
\caption{Impact of the number of antennas in the considered multi-waveguide pinching-antenna system, where $D_y=8$~m, $N=8$, $P_t=10$~dBm, and $R_\mathrm{min} = 0.1$~bps/Hz.}
\label{result4}
\end{figure}

The impact of the number of available pinching antennas is presented in \fref{result4}. For the case where $K=3$, the waveguides are positioned at $y_k^\mathrm{Pin}=0$ and $y_k^\mathrm{Pin}=\pm 2.6667$. As the number of pinching antennas increases, the positions of the activated antennas can be adjusted more precisely. This leads to an increase in both the sum rate and the number of activated antennas. Moreover, the deployment of additional waveguides enhances the sum rate and facilitates the activation of a greater number of pinching antennas, and the SCA based power allocation algorithm also contributes to sum rate improvement across various scenarios. \fref{result4} further demonstrates that the proposed optimal SIC decoding order can achieve a higher sum rate with fewer activated antennas, thereby improving overall system efficiency.

\begin{figure}[!t]
\centering{\includegraphics[width=84mm]{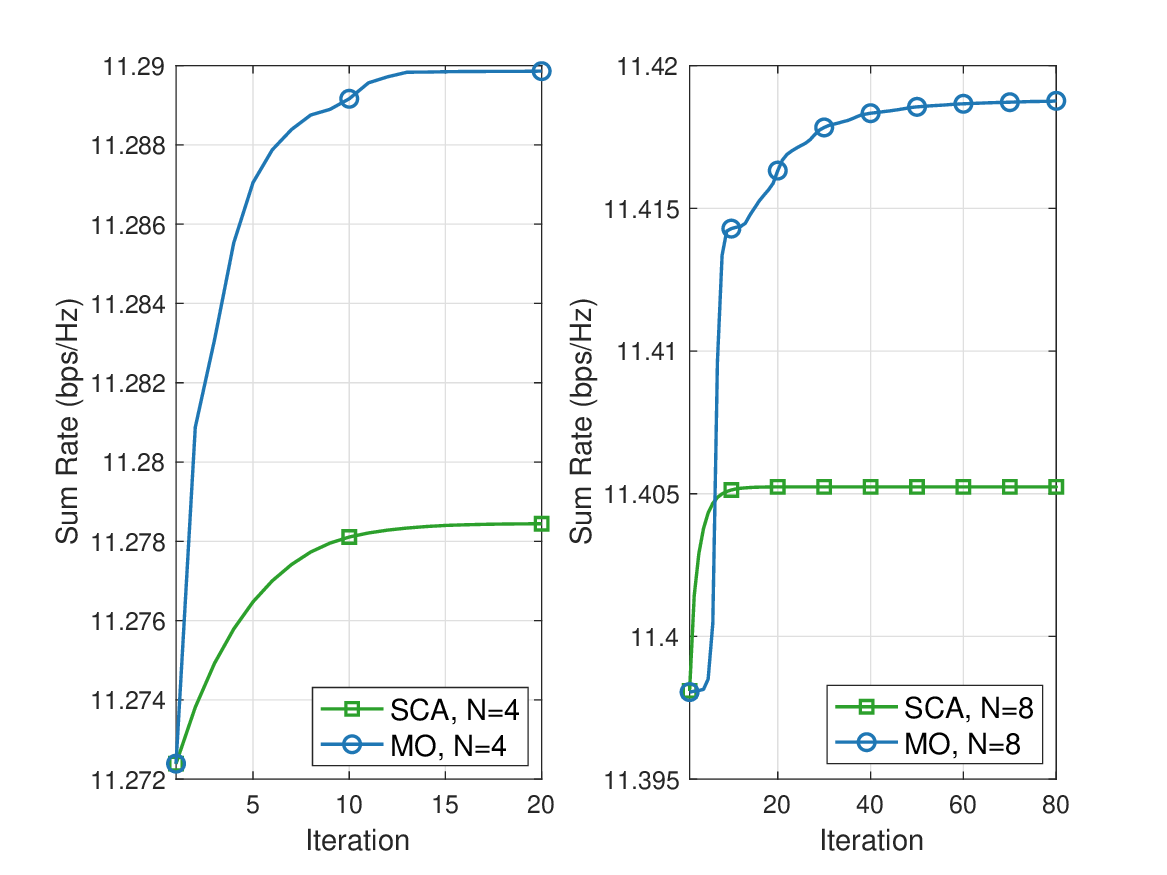}}
\caption{Convergence performance of the proposed power allocation solutions, where $D_y=8$~m, $K=1$, $M=20$, $P_t=10$~dBm, and $R_\mathrm{min} = 0.1$~bps/Hz.}
\label{result5}
\end{figure}

The convergence behavior of the proposed power allocation solutions are presented in \fref{result5}, where each vertex selection in the polyblock outer approximation algorithm is considered as an iteration of the monotonic optimization based solution (denoted by MO in the figure). In general, the SCA based power allocation approach can achieve a near-optimal solution with substantially lower computational complexity. In contrast, the monotonic optimization based method converges to a higher sum rate but typically requires more iterations. In the case that the number of users is large, the convergence of the monotonic optimization based algorithm becomes noticeably slower, highlighting its scalability limitations in scenarios with a large number of users.

\begin{figure}[!t]
\centering{
\subfigure[$K=2$]{\centering{\includegraphics[width=84mm]{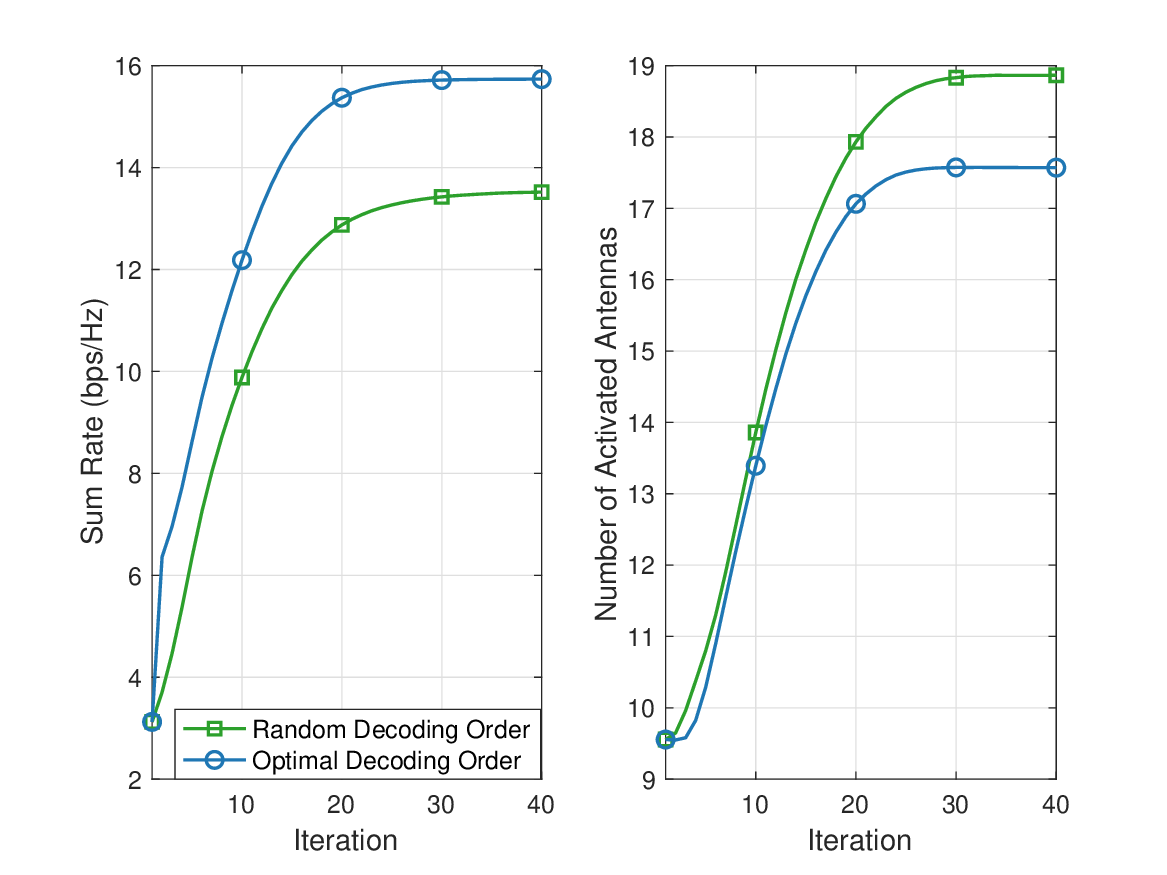}}}
\subfigure[$K=3$]{\centering{\includegraphics[width=84mm]{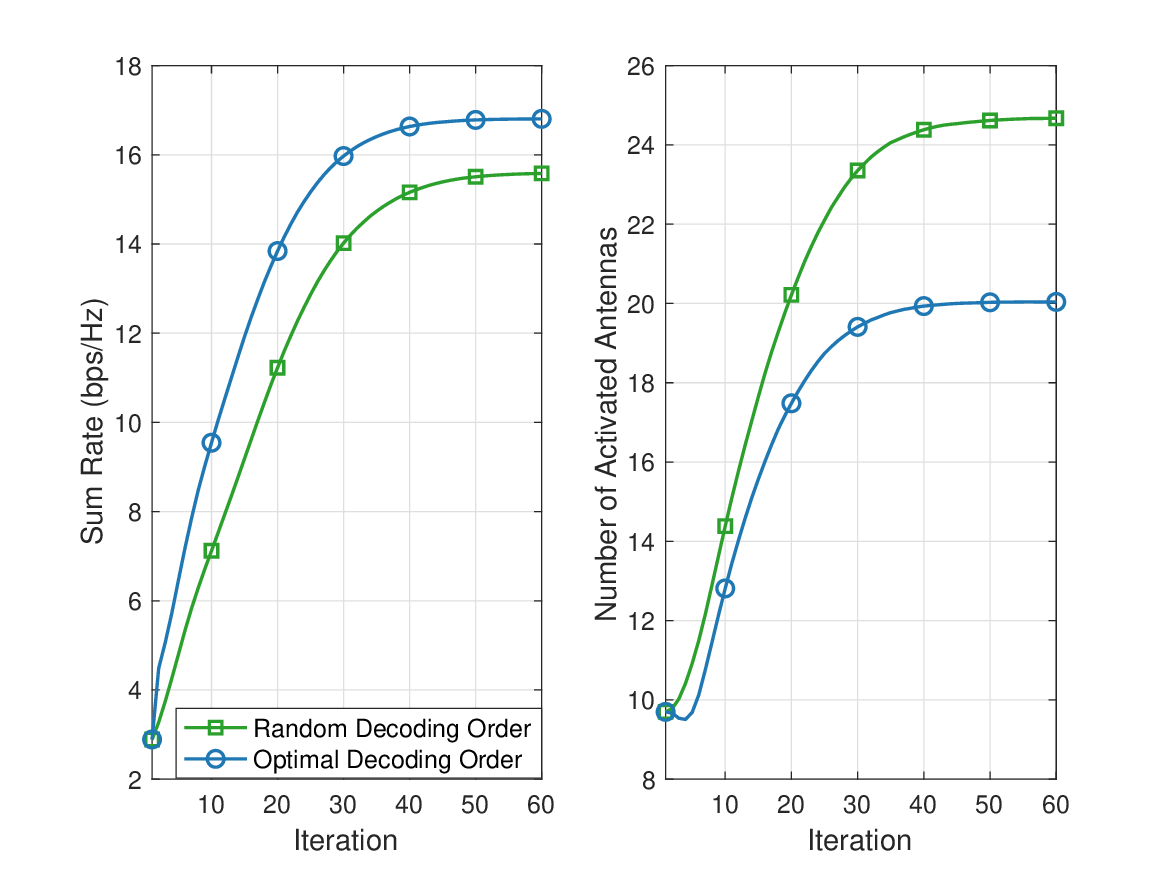}}}}
\caption{Convergence performance of the proposed coalitional game based algorithm, where $D_y=8$~m, $M=50$, $N=10$, and $P_t=10$~dBm.}
\label{result6}
\end{figure}

\fref{result6} presents the convergence behavior of the proposal coalitional game based waveguide assignment and antenna activation algorithm for different SIC decoding orders. It can be observed that, with the optimal SIC decoding order, a higher sum rate is achieved with fewer activated pinching antennas, and the number of required iterations for convergence is reduced. Additionally, in the case that three waveguides are implemented, since the number of potential pinching antennas  increases, additional iterations for convergence should be performed, leading to a further improvement in the achievable sum rate.
%%%%%%%%%%%%%%%%%%%%%%%%%%%%%%%%%%%%%%%%%%%%%%%%%
%%%%%%%%%%%%%%%%%%%%%%%%%%%%%%%%%%%%%%%%%%%%%%%%%
\section{Conclusions}
In this paper, a multi-waveguide pinching-antenna system was studied in a practical scenario where pre-installed pinching antennas were activated/deactivated to serve multiple users in a NOMA manner. To improve the spectral efficiency, a sum rate maximization problem was formulated by jointly considering waveguide assignment, antenna activation, power allocation, and SIC decoding order design. The formulated problem was decoupled into two sub-problems, where multiple approaches were used to effectively obtain the solutions. Specifically, by deriving the optimal SIC decoding order, a coalitional game based algorithm was developed for waveguide assignment and antenna activation. For the power allocation problem, monotonic optimization and SCA were employed to obtain globally optimal and low-complexity solutions, respectively. Simulation results demonstrate that the proposed NOMA-based multi-waveguide pinching-antenna system exhibits superior performance compared to both OMA-based and single-waveguide pinching-antenna systems. Furthermore, the proposed solutions have the potential to significantly improve the sum rate and reduce the outage probability.
%%%%%%%%%%%%%%%%%%%%%%%%%%%%%%%%%%%%%%%%%%%%%%%%%
%%%%%%%%%%%%%%%%%%%%%%%%%%%%%%%%%%%%%%%%%%%%%%%%%
%%%%%%%%%%%%%%%%%%%%%%%%%%%%%%%%%%%%%%%%%%%%%%%%%
\section*{Appendix~A: Proof of Proposition~\ref{order}}
\eqr{decoderate} and \eqr{achirate} indicate that on any waveguide $k$, the achievable data rate of any assigned user $n$ is affected by the users that follow it in the decoding order $Q_k$, as follows:
\begin{align}\nonumber
R_{k,n}^{(\mathrm{Q}_k)}&=\!\min\!\left\{\!\log_2\!\!\left(\!\!1\!\!+\!\!\frac{p_{k,n}\frac{P_t}{M_k}|h_{k,n'}|^2}{\frac{P_t}{M_k}|h_{k,n'}|^2\!\sum_{i=n\!+\!1}^N\!\alpha_{k,i}p_{k,i}\!\!+\!\!I_{k,n'}\!\!+\!\!\sigma^2}\!\!\!\right) \!\!\Bigg|\forall n'\!\ge\! n\!\!\right\}\\
&=\min\!\left\{\!\log_2\!\!\left(\!\!1\!\!+\!\!\frac{p_{k,n}\frac{P_t}{M_k}}{\frac{P_t}{M_k}\!\sum_{i=n+1}^N\!\alpha_{k,i}p_{k,i}\!+\!\!\frac{I_{k,n'}\!+\sigma^2}{|h_{k,n'}|^2}}\!\right)\!\Bigg|\forall n'\!\ge\! n\right\}.
\end{align}
This equation can be equivalently transformed into
\begin{equation}
R_{k,n}^{(\mathrm{Q}_k)}\!\!\!=\!\!\log_2\!\!\!\left(\!\!1\!\!+\!\!\frac{p_{k,n}\frac{P_t}{M_k}}{\frac{P_t}{M_k}\!\!\sum_{i=n+1}^N\!\alpha_{k,i}p_{k,i}\!+\!\max\!\left\{\!\frac{I_{k,n'}\!+\sigma^2}{|h_{k,n'}|^2}\!\Big|\forall n'\!\ge\! n\!\right\}}\!\!\right).
\end{equation}
It can be concluded from the above equation that in order to maximize the achievable data rate of user $n$, the impact caused by the subsequent users in decoding order $Q_k$ should be eliminated. In other words, the users assigned to waveguide $k$ should be sorted based on the following order:
\begin{equation}
\max\!\left\{\frac{I_{k,n'}\!\!+\!\sigma^2}{|h_{k,n'}|^2}\Big|\forall n'\!\ge\! 1\right\}\ge\max\!\left\{\frac{I_{k,n'}\!\!+\!\sigma^2}{|h_{k,n'}|^2}\Big|\forall n'\!\ge\! 2\right\}\ge\cdots.
\end{equation}
This condition can be simplified as follows:
\begin{equation}
\frac{I_{k,1}+\sigma^2}{|h_{k,1}|^2}\ge \frac{I_{k,2}+\sigma^2}{|h_{k,2}|^2}\ge\cdots.
\end{equation}
By making the newly introduced subscripts satisfy the above order, this proposition can be proved.\QEDA
%%%%%%%%%%%%%%%%%%%%%%%%%%%%%%%%%%%%%%%%%%%%%%%%%
%%%%%%%%%%%%%%%%%%%%%%%%%%%%%%%%%%%%%%%%%%%%%%%%%
\bibliographystyle{IEEEtran}
\bibliography{KaidisBib}
%%%%%%%%%%%%%%%%%%%%%%%%%%%%%%%%%%%%%%%%%%%%%%%%%
%%%%%%%%%%%%%%%%%%%%%%%%%%%%%%%%%%%%%%%%%%%%%%%%%
\end{document}